\documentclass[conference,compsoc]{IEEEtran}
\usepackage{cite}
\usepackage{amsmath,amssymb,amsfonts}
\usepackage{graphicx, subcaption}
\usepackage{textcomp}
\usepackage{xcolor}\usepackage{mdframed}
\usepackage{graphicx}
\usepackage{algorithm}
\usepackage{listings}
\usepackage{gensymb}
\usepackage[utf8]{inputenc}
\DeclareUnicodeCharacter{2212}{-}
\usepackage{mathtools}
\usepackage{algpseudocode}
\usepackage{multirow}
\usepackage{amssymb}
\usepackage{url}
\usepackage{hyperref}
\usepackage{xcolor,colortbl}
\usepackage{array}
\usepackage{booktabs}
\usepackage{pifont}
\usepackage{verbatim}
\usepackage{tikz}

\usepackage{makecell}

\usepackage{amsmath, amssymb, amsfonts}
\usepackage{enumitem}
\usepackage{siunitx}
\usepackage{xspace}
\usepackage{listings}
\usepackage{xcolor}

\usepackage[export]{adjustbox}

\definecolor{mGreen}{rgb}{0,0.6,0}
\definecolor{mGray}{rgb}{0.5,0.5,0.5}
\definecolor{mPurple}{rgb}{0.58,0,0.82}
\definecolor{backgroundColour}{rgb}{0.95,0.95,0.92}
\definecolor{darkpastelred}{rgb}{0.76, 0.23, 0.13}

\lstdefinestyle{customc}{
  belowcaptionskip=1\baselineskip,
  moredelim=**[is][\color{red}]{@}{@},
  breaklines=true,
  frame=L,
  numbers=left,
  xleftmargin=13pt,
  numbersep=5pt, 
  language=C++,
  showstringspaces=false,
  basicstyle=\footnotesize\ttfamily,
  keywordstyle=[1]\bfseries\color{blue!40!black},
  commentstyle=\itshape\color{purple!40!black},
  identifierstyle=\color{black},
  stringstyle=\color{orange},  
  keywords=[2]{malloc},
  keywordstyle=[2]\bfseries\color{darkpastelred},
  morekeywords={munmap},
}
\renewcommand{\paragraph}[1]{\medskip \noindent {\bf #1.}}
\newcommand{\hl}[1]{\textcolor{black}{#1}}

\usepackage{pifont}
\newcommand{\cmark}{\textcolor{green}{\ding{51}}}%
\newcommand{\xmark}{\textcolor{red}{\ding{55}}}%
\newcommand{\etal}{\textit{et al.}}
\newcommand*\blackcircled[1]{\tikz[baseline=(char.base)]{
            \node[shape=circle, draw, inner sep=1pt, fill=black, text=white, font=\sffamily\scriptsize] (char) {#1};}}
\newcommand{\attack}{\textsc{Fault$+$Probe}\xspace}

\def\BibTeX{{\rm B\kern-.05em{\sc i\kern-.025em b}\kern-.08em
    T\kern-.1667em\lower.7ex\hbox{E}\kern-.125emX}}
\begin{document}

\title{\attack: A Generic Rowhammer-based Bit Recovery Attack}

\author{\IEEEauthorblockN{Kemal Derya, M. Caner Tol, and Berk Sunar}
\IEEEauthorblockA{Worcester Polytechnic Institute, Worcester, MA, USA\\
\{kderya, mtol, sunar\}@wpi.edu}
}

\maketitle


\begin{abstract}

Rowhammer is a security vulnerability that allows unauthorized attackers to induce errors within DRAM cells, e.g., to attain elevated user privileges or to extract sensitive information from cryptographic schemes. To prevent fault injections from escalating to successful attacks, a widely accepted mitigation is implementing fault checks on instructions and data. 

We challenge the validity of this assumption by examining the impact of the fault on the victim's functionality, as opposed to solely focusing on the victim's erroneous output. Specifically, we illustrate that an attacker can construct a profile of the victim's memory based on the directional patterns of bit flips. This profile is then utilized to identify the most susceptible bit locations within DRAM rows. These locations are then subsequently leveraged during an online attack phase with side information observed from the change in the victim's behavior to deduce sensitive bit values. Consequently, the primary objective of this study is to utilize Rowhammer as a probe, shifting the emphasis away from the victim's memory integrity and toward statistical fault analysis (SFA) based on the victim's operational behavior.

In the signing victim attack scenario, we show \attack\ may be used to circumvent the verify-after-sign fault check mechanism, which is designed to prevent the generation of erroneous signatures that leak sensitive information. It does so by injecting directional faults into key positions identified during a memory profiling stage. The attacker observes the signature generation rate and decodes the secret bit value accordingly. This circumvention is enabled by an observable channel in the victim. \attack\ is not limited to signing victims and can be used to probe secret bits on arbitrary systems where an observable channel is present that leaks the result of the fault injection attempt. To demonstrate the attack, we target the fault-protected ECDSA in wolfSSL's implementation of the TLS 1.3 handshake. We recover 256-bit session keys with an average recovery rate of 22 key bits/hour and a 100\% success rate.

\end{abstract}


\section{Introduction}
\paragraph{Rowhammer} The discovery of Rowhammer by Kim \etal ~\cite{kim2014flipping} led to a number of attacks, such as gaining kernel privileges by corrupting Page Table Entries~\cite{seaborn2015exploiting}, bypassing authentication using opcode flipping~\cite{gruss2018another}. Rowhammer has been shown to be a threat in cloud environments~\cite{xiao2016one, cojocar2020susceptible}, mobile platforms \cite{van2016drammer} and, network connections\cite{tatar2018throwhammer,lipp2020nethammer}. Rowhammer has also been shown to work from Javascript~\cite{gruss2016rowhammerjs,deridder2021smash}, extending the attack surface to include browsers. 
To mitigate Rowhammer, detection-based~\cite{irazoqui2016mascat,chiappetta2016real, zhang2016cloudradar, herath2015these, payer2016hexpads, gruss2016flush+, aweke2016anvil, corbet2016kernel}, and prevention based~\cite{gruss2016rowhammerjs,van2016drammer,brasser2017cant} countermeasures have been proposed. Gruss \etal~\cite{gruss2018another} showed that each can be defeated. Similarly, row-refresh-based mitigation attempts implemented in DDR4 chips have been easily bypassed ~\cite{frigo2020trrespass,jattke2022blacksmith,kogler2022halfdouble}. Cojocar \etal~\cite{cojocar2019ecc} showed how to induce bit flips on Error-correcting-code (ECC) memories using the timing side channels.

\paragraph{Attacks on Signature Schemes}
Beyond system-level attacks, cryptographic schemes have also been more directly targeted using Rowhammer. For instance, \cite{mus2020quantumhammer} uses Rowhammer faults to recover key bits in the LUOV post-quantum (PQ) signature scheme (a Round 2 competitor in the PQ NIST standard competition). The technique was formalized and named \textit{Signature Correction Attack} (SCA) in \cite{islam2022dilithium}, which targeted CRYSTALS-Dilithium (PQ NIST Standard). In a nutshell, SCA  introduces faults using Rowhammer while signatures are being computed, and by mathematically tracing and correcting the faulty signatures, it is able to recover key bit values. Later, the same attack with modification of the correction mechanisms was shown to work in traditional signatures schemes, e.g., ECDSA, EdDSA, and RSA signatures in~\cite{mus2023jolt}. The SCA on Dilithium was further improved by using (non-Rowhammer) instruction skip attacks in \cite{DilithimSCA24}. 

\paragraph{Fault Analysis Methods}
The fault's effect on the victim has been examined in various scenarios. Safe-error Analysis (SEA) was introduced in~\cite{yen2000checking} to show that fault checks on the output create another side channel to be exploited. Later, Ineffective Fault Analysis (IFA) was used in~\cite{clavier2007secret}, which shows the ineffectiveness of the counter-measure based on releasing the output if the results from executing the victim twice are equal. Both SEA and IFA are based on induced faults that do not alter the output. On the other hand, the faulty outputs are utilized to recover secret keys. Biham and Shamir~\cite{biham1997differential} used Differential Fault Analysis (DFA) on DES to extract the secret key from the faulty ciphertexts. Statistical Fault Analysis (SFA) was introduced in~\cite{fuhr2013fault}, where the AES key is recovered without the knowledge of input messages. Furthermore, Statistical Ineffective Faulty Analysis (SIFA)~\cite{dobraunig2018sifa} was introduced to generalize SFA and IFA. Yet, none of the prior works could show effectiveness in a local attack scenario where physical access is not possible.

\paragraph{Library Countermeasures}
To mitigate Rowhammer-enabled fault injection attacks on crypto schemes, e.g.~\cite{mus2020quantumhammer,islam2022dilithium,mus2023jolt,FrodoKemattack,LWERowhammer}, most authors recommend hardware mitigations, e.g. increased refresh rates. Further software-based fault detection techniques are recommended, i.e., SCA papers~\cite{mus2020quantumhammer,islam2022dilithium,mus2023jolt} advocate checking faulty signatures before releasing them to potential adversaries. An adversary cannot run the signature correction step without access to the faulty signatures. 
As for industry response, due to the lack of effective and efficient hardware mitigations in DRAM chips, library developers started to patch their own software with application-specific hardening mechanisms, e.g., \texttt{SUDO} developers implemented a logic hardening that prevents authentication bypass~\cite{adiletta2023mayhem} and wolfSSL developers implemented a \textit{verify-after-sign} check to mitigate signature-correction attacks~\cite{mus2023jolt}. 
In light of the recent advances in Rowhammer attacks, we pose the following questions:
\begin{itemize}[nosep,leftmargin=*]
\item \textit{Are there any observable channels when combined with Rowhammer faults that might be exploited to recover secrets?}
\item \textit{If granted, how can we use such channels along with Rowhammer to probe and recover secret bits from the victim's memory?}
\item \textit{Can such an attack be used to bypass existing detection-based defenses?}
\end{itemize}

\subsection{Our Contributions}

This work introduces a generic technique that enables attackers to probe bit values in a victim's memory space using Rowhammer faults. It uses SFA to exploit the behavioral changes in the victim's execution during fault injection. These behavioral changes confirm a successful fault being injected (or otherwise) and provide a feedback mechanism. 

\paragraph{Challenges} To use this side channel for an end-to-end attack, one needs to overcome the following challenges:
\begin{itemize}[nosep,leftmargin=*]
    \item Finding targets with observable feedback mechanisms when a fault is introduced on the secret value,
    \item Finding physical pages in the memory that produce repeatable flips in a target offset yet do not produce too many flips in other offset locations,
    \item Mapping the victim to target flippy page and hammering,
    \item Amplifying the correct prediction probability of the secret bit by reclaiming the same flippy page and repeating the attack.
    
\end{itemize}
\paragraph{Contributions}
To solve the challenges we explained,
\begin{itemize}[noitemsep,leftmargin=*]
\item We introduce \attack, a novel fault-based side-channel leakage mechanism that utilizes SFA on a victim program caused by Rowhammer faults to recover secret bits on co-located platforms. (Section~\ref{sec:attack})

\item We introduce three kinds of behavioral changes that serve as feedback mechanisms  observable by the adversary:
\begin{enumerate}[nosep,leftmargin=*]
    \item Fault-checking mechanisms that were put in place to prevent Rowhammer from succeeding may become leakage channels. For example, a signature generation scheme naturally slows down if fault detection is implemented and faults are injected during the signing. The adversary can measure the signing rate to deduce whether fault injection attempts succeed or not.
    
    \item Many protocols report failures to assist diagnosis. For instance, wolfSSL's TLS implementation returns connection error codes, providing information about a fault's success, which the attacker might exploit with greater accuracy.
    
    \item Finally, without any error code or slowdown in the signing rate, the adversary can potentially receive a faulty signature, which fails to be verified by the public verification algorithm, indicating a fault on the victim's side to the adversary\footnote{Unlike in Signature Correction Attacks, \attack\ does not need to correct the signature but only runs verify primitive to determine the success or failure of the fault.}.
\end{enumerate}

\item We provide extensive experimental evidence using a thorough offline profiling phase on DDR4 memory devices that identifies memory pages with stable but isolated directional flips. We show that certain memory locations in DRAM modules have low-noise and repeatable flips, which enables \attack. (Section~\ref{sec:offline_profiling})

\item We propose a novel method to reclaim flippy pages that enables us to reuse them multiple times for hammering a secret value. (Section~\ref{sec:reclaiming}) 

\item  We demonstrate an end-to-end attack on a TLS 1.3 handshake protocol implemented in wolfSSL. We use wolfSSL's ECDSA implementation, which is specifically protected against fault injections with the \textit{verify-after-sign} check mechanism. \hl{We show that \attack can recover 256-bit ECDSA private key from wolfSSL with an average recovery rate of 22 bits/hour and 100\% success rate.} (Section~\ref{sec:wolfssl})
\end{itemize}

\section{Background}
\begin{figure}%
    \centering
    \subfloat[\centering Double-sided \label{fig:a1}]{{\includegraphics[width=0.20\linewidth]{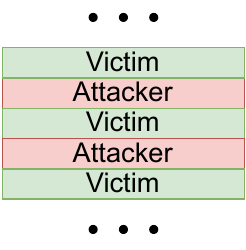}}}%
    \hfill
    \subfloat[\centering N-sided \label{fig:b1}]{{\includegraphics[width=0.20\linewidth]{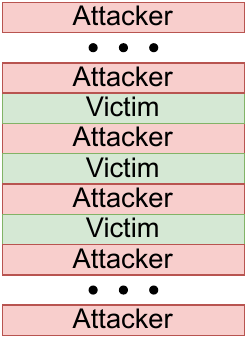}}}%
    \hfill
    \subfloat[\centering N-sided multi-bank \label{fig:c1}]{{\includegraphics[width=0.35\linewidth]{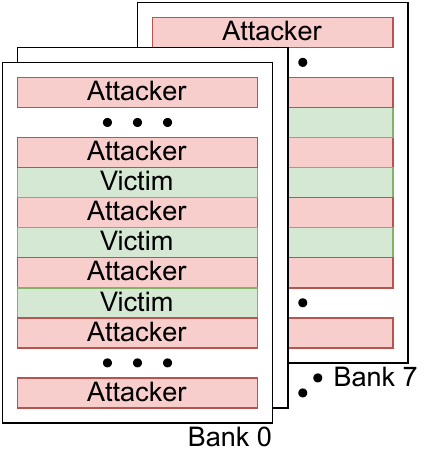}}}%
    \caption{Different Rowhammer hammering configurations}%
    \label{fig:rowh}%
\end{figure}

\paragraph{Rowhammer}
%
Fig.~\ref{fig:rowh} displays the arrangement of memory in different Rowhammer setups, in which each DRAM row comprises two pages of 4KiB each. In these configurations, it is presumed that the attacker can manipulate pages containing known data, referred to as attacker pages. The rows designated for the victim hold secret data from the victim's process. As depicted in Fig.~\ref{fig:a1} and~\ref{fig:b1}, all rows are located within the same bank. However, in Fig~\ref{fig:c1}, rows are allocated on different banks. The attacker and victim pages are positioned adjacently. Consequently, frequent accesses to the attacker pages can cause bit flips in the victim pages. 

Previous work introduced the building blocks of an end-to-end Rowhammer attack, such as finding contiguous physical memory~\cite{islam2019spoiler, kangsledgehammer24}, and memory massaging~\cite{chakraborty2020explframe,kwong2020rambleed} which enabled more precise Rowhammer attacks~\cite{tatar2018hammertime}.

Weissman \etal~\cite{weissman2019jackhammer} showed up to three times faster attack times in heterogeneous FPGA-CPU platforms when Rowhammer is launched from the FPGA. Tobah \etal~\cite{tobah2022spechammer} showed Rowhammer can increase the number of potential Spectre-V1~\cite{Kocher2018spectre} gadgets by 200 times. Recently, Kang \etal~\cite{kangsledgehammer24} introduced a multi-bank hammering method that increases the throughput of bit flips up to seven times. Adiletta \etal~\cite{adiletta2023mayhem} demonstrated that Rowhammer can also be used to corrupt register values temporarily stored in memory.

\paragraph{Statistical Fault Analysis}
SEA~\cite{yen2000checking} uses faults that do not change the outputs. It injects faults on the intermediate values and observes if the output is changed to deduce the secret bits. SIFA~\cite{dobraunig2018sifa} uses principles of IFA and SFA. It utilizes the non-uniform probability of ineffective faults that result in errors. Moreover, Fuhr \etal~\cite{fuhr2013fault} used SFA to recover the secret key from AES. The method works by injecting fault to an intermediate value. Every faulty ciphertext is partially decrypted for each key candidate. Then, the key candidates are statistically evaluated to isolate the key.

\paragraph{Mitigation Techniques}
Detection~\cite{irazoqui2016mascat,chiappetta2016real, zhang2016cloudradar, herath2015these, payer2016hexpads, gruss2016flush+, aweke2016anvil, corbet2016kernel}, and prevention based~\cite{gruss2016rowhammerjs,van2016drammer,brasser2017cant} countermeasures have been proposed to mitigate Rowhammer attacks. Yet, Gruss \etal~\cite{gruss2018another} showed that each can be defeated. Cojocar \etal~\cite{cojocar2019ecc} showed how to induce bit flips on Error-correcting-code (ECC) memories using the timing side channels. Similarly, row-refresh-based mitigation attempts implemented in DDR4 chips have been easily bypassed by adopting different hammering patterns~\cite{frigo2020trrespass,jattke2022blacksmith,kogler2022halfdouble}. Recently proposed mitigations, such as~\cite{wang2021discreetpara, juffinger2023csi,marazzi2023rega}, require hardware design changes, and DRAM vendors have not yet adopted them. Infection-based countermeasures~\cite{gierlichs2012infective,tupsamudre2014destroying} have been proposed to make the faulty ciphertexts useless to mitigate SFA.

\section{Threat Model}
Following the prior work on Rowhammer-based attacks~\cite{kim2014flipping,gruss2018another,gruss2016rowhammerjs,xiao2016one,cojocar2020susceptible,mus2023jolt,mus2020quantumhammer} and microarchitectural side-channel attacks~\cite{canella2019fallout,Kocher2018spectre,Lipp2018meltdown,vanbulck2020lvi,vanschaik2019ridl}, attacker and victim are co-located in the same system in our threat model. The adversary has no physical access to the processor or memory system but can run code on the target machine. We assume the system is free of any software vulnerability and all the protections against microarchitectural side-channel attacks are deployed.  We assume the attacker can access a mechanism that tells if the secret has been corrupted. Such a mechanism can naturally exist in different applications without any attacker effort. For instance, the signature verification algorithm in public key crypto schemes can serve as an oracle if a fault was injected on secret or not~\cite{mus2023jolt}. If the victim does not reveal faulty signatures, the error code returned to the attacker can do the same task. In extreme cases where the victim does not return any error code and retries the correct execution, the timing difference caused by extra execution can reveal if the fault was injected successfully.

\section{\attack Attack}\label{sec:attack}

Earlier Rowhammer attacks depended on introducing faults into the victim's process. The attacks in ~\cite{mus2020quantumhammer, islam2022dilithium, mus2023jolt}, i.e. applied correction techniques to faulty values, extracting secret bits from the victim's side. Mitigation of these attacks has been attempted through check mechanisms such as the \textit{verify-after-sign} method, aiming to prevent the generation of faulty signatures~\cite{wolfssl}. Nevertheless, this work demonstrates that the \attack attack can still access secret bits even when such check mechanisms are in place. This enables unprivileged attackers to read secret bits by monitoring the flip rate of Rowhammer-induced bit flips at specific memory locations. Since it needs to find a correlation between the profile of the victim's memory and the number of changes observed in the victim's functionality, \attack may be viewed as a method that utilizes the principles of SFA and SEA.

\subsection{Attack Overview}

\begin{figure*}[h!]
    \centering
    \includegraphics[width=\linewidth]{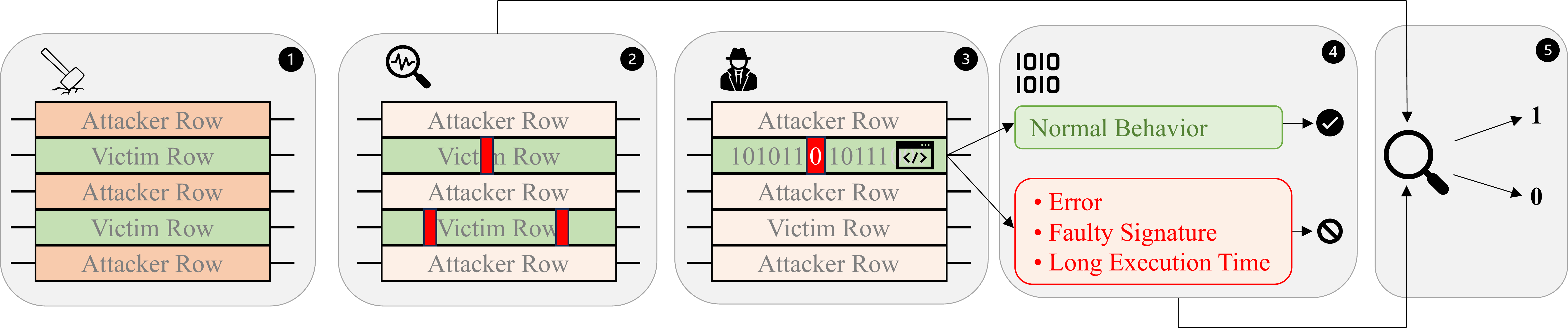}
    \caption{\attack Attack overview. \protect\blackcircled{1} The attack starts by hammering adjacent rows for the offline memory profiling phase. \protect\blackcircled{2} The most sensitive bit offsets are profiled. \protect\blackcircled{3} The victim process is located on the profiled row. \protect\blackcircled{4} The process is run and hammered during the online attack phase a sufficient number of times. \protect\blackcircled{5} The value of the targeted offset is found by comparing the results from the offline memory profiling and online attack phases.}
    \label{fig:attack}
\end{figure*}

The \attack attack exploits the reproducibility of bit flips on DRAM rows through meticulous profiling. The secret bits are probed by repetitively flipping them. Fig.~\ref{fig:attack} shows the attack overview. Our attack is divided into the steps below.

    \noindent\blackcircled{1} We find a contiguous memory block and hammer it to find flippy bit locations. This stage is referred to as \textit{offline memory profiling}. This stage is performed until a sufficient number of victim pages for the online attack is found. (Section~\ref{sec:offline_profiling})
    
   \noindent\blackcircled{2} We analyze the victim pages to observe the directional bit flips. We construct a profile based on the bit flips. The most sensitive bit locations within victim pages are identified. (Section~\ref{subsec:profiling})
   
    \noindent\blackcircled{3} We choose a victim page from the profile. This victim page has reproducible bit flips on specific bit locations. Before going into the online attack phase, we search the memory space for the victim page with its corresponding attacker pages. We allocate dummy pages to place the secret bits into the victim page for the online attack phase. We manipulate the Linux Buddy Allocator and unmap the dummy pages to map the secret bits to the victim page~\cite{kwong2020rambleed, mus2023jolt}. We target the secret bits using Rowhammer for the online attack phase. (Section~\ref{sec:online_attack})
    
    \noindent\blackcircled{4} We look at the changes in the victim's behavior. After repeatedly running the victim process and hammering the victim page, we record the observable results. (Section~\ref{sec:online_attack})
    
    \noindent\blackcircled{5} We compare the victim page's profile with the operational behavior of the victim. Then, we probabilistically determine the original secret bit.  (Section~\ref{sec:online_attack})

\paragraph{Reproducible Bit Flips} Reproducible bit flips refer to the ability to consistently induce specific changes in memory content through controlled Rowhammer attacks. By carefully targeting and repeatedly accessing specific rows of DRAM, attackers can trigger predictable bit flips, potentially leading to unauthorized access or system compromise~\cite{adiletta2023mayhem}. Our approach achieves as low as single-bit reproducible flips within the same row. We fine-tune the attack parameters to minimize the impact on neighboring bits. This work shows the precision and control of Rowhammer attacks for improved bit probing.

\subsection{Utilizing Reproducible Bit Flips}

This section details the method for utilizing reproducible bit flips to indirectly infer the contents of the victim's bits, bypassing the need to mathematically trace the fault directly.

\paragraph{Activating Attacker Pages} Executing a hammering routine on pages controlled by the attacker leads to interference within the victim's pages. \hl{The hammering routine plays an important role in reproducible bit flips. The number of bit flips is affected by the number of attacker pages. Moreover, we observe that the position of the victim page among the attacker pages affects the number of bit flips on the victim page. Even the data stored in the attacker pages affects the distribution of bit flips on the victim page. Therefore, we need to activate the attacker pages by using a hammering routine that generates precise and reproducible bit flips.}

\paragraph{Deducing Secret Bits} Prior to the attack, victim pages undergo a profiling stage to ascertain the flipping tendencies of individual bits. A victim page that shows precise and reliable bit flips is chosen from the profiling stage. Then, the online attack is started by locating the victim process on the victim page, which is hammered repetitively. 

The attack's efficacy is monitored by examining the disturbance in the victim's processes. Given that bit flips do not occur with every attempt, there is a calculable probability associated with the attack's success. Concurrence between the outcomes of the profiling phase and the online attack iterations allows for the inference of the probed bit.

\subsection{Offline Memory Profiling} \label{sec:offline_profiling}

This section describes the memory profiling techniques to perform \attack attack explained in Section \ref{sec:attack}.

\begin{figure*}[h!]
    \centering
    \includegraphics[width=\linewidth]{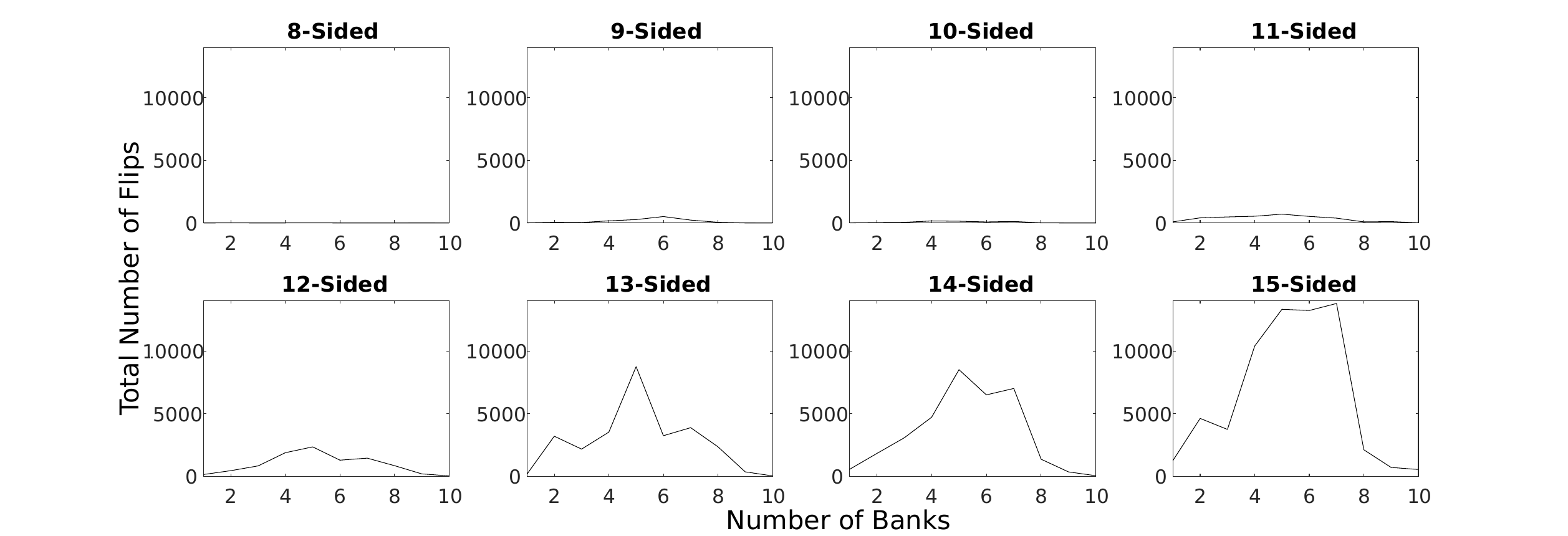}
    \caption{Optimal configuration search for the most number of bit flips.}
    \label{fig:flipvsbanks}
\end{figure*}

\subsubsection{Finding Physically Contiguous Memory Pages}

Our approach starts by identifying contiguous memory addresses to position attacker rows near the victim rows. One technique to find contiguous physical memory involves using huge pages, ensuring that sequential virtual addresses correlate directly with contiguous physical addresses. However, this method necessitates unique system configurations. Operating under the constraint of standard configurations, we leverage DRAMA~\cite{pessl2016drama} on DRAM modules. For an in-depth understanding of the DRAMA exploit, we direct readers to the dedicated section detailing its implementation. Other alternatives for finding contiguous physical memory segments from user space include~\cite{islam2019spoiler} and \cite{kangsledgehammer24}. 


\subsubsection{Finding Adjacent Rows on DRAM}

We start by allocating a memory chunk from the system. On this chunk of memory, we search for the memory space that is located on the same bank. Depending on the number of banks we need to allocate, we gather each memory space associated with its bank into a memory array. The memory array contains the physical addresses of memory chunks located in different banks. We use the method presented in~\cite{pessl2016drama} to find the memory chunks located in the same bank. 

Once memory blocks on different banks have been allocated, our next step involves identifying adjacent pages within the same DRAM bank. To achieve this, we employ the method in~\cite{pessl2016drama}, which reveals the row number of physical addresses in memory chunks on different banks. In each bank, we look for physical addresses that are either located in the same row or the next row. The number of adjacent rows might be different for each bank. We truncate the number of adjacent rows to the minimum number of rows found in a bank. If we do not find any adjacent rows in the same bank, we start the process from the beginning by mapping a memory chunk from the system.

\subsubsection{Profiling the Contiguous Memory}
\label{subsec:profiling}

Upon successfully arranging rows within the same DRAM bank of a contiguous physical memory block, we executed an effective Rowhammer attack. In systems equipped with DDR3 DRAM, we employed a double-sided Rowhammer technique, hammering each victim row with an attacker row on both sides. This approach, however, is rendered ineffective on DDR4 systems due to Target Row Refresh (TRR) countermeasures. Consequently, we adapted to an N-sided Rowhammer strategy~\cite{frigo2020trrespass,jattke2022blacksmith,kangsledgehammer24}, aiming at multiple victim rows with a sequential pattern of attacker and victim rows. Subsequently, we pinpointed the rows experiencing bit flips by persistently hammering the attacker rows and monitoring for changes in the victim rows' data.

\paragraph{Multi-bank Hammering} The number of banks used in hammering directly impacts the number of flips observed. Our observations reveal that employing a multi-bank approach generates more bit flips than a single bank. Kang~\etal~\cite{kangsledgehammer24} introduced the multi-bank hammering method. We use their method to observe the effect of multi-bank hammering. We utilize Linux's huge pages to target the same DRAM pages for a more fair comparison between the configurations. We need to target the same pages so that we would have more understanding of the number of bit flips with different numbers of attacker rows and different numbers of banks. 
Fig.~\ref{fig:flipvsbanks} shows the total number of bit flips with different numbers of attacker rows and multi-bank hammering on the same DRAM pages.
The number of bit flips refers to the number of total flips from 1 $\rightarrow$ 0 and 0 $\rightarrow$ 1 flips.
Fig.~\ref{fig:flipvsbanks} indicates that increasing the number of banks increases the number of flips observed until eight banks in most cases. It decreases dramatically after using more than eight banks. 

\paragraph{Memory Profiling} \hl{We set the attack window, which is dictated by the number of rows and the number of banks. We sweep contiguous chunks of memory on each bank by using the attack window, which contains the attacker and victim rows. Firstly, we populate the victim rows with 0s and the attacker rows with 1s. Then, we hammer the attacker rows by 500K times to observe the bit flips on the victim rows. Secondly, we swap the values of attacker and victim rows and hammer them to observe the bit flips. If we find any bit flips on the victim rows, we use them for memory profiling. If we do not observe any bit flips during this phase, we pass to the next attack window on the contiguous memory chunk on each bank.}

The data pattern within the victim row influences the bit flip rate at specific positions. Typically, bit configurations of 1-0-1 and 0-1-0 are more prone to flips than those of 1-1-1 and 0-0-0~\cite{kim2014flipping} when the central bit is the victim bit. 

\hl{To simulate realistic flip rates, the victim rows are filled with random data during the profiling phase. The process begins by setting the attacker rows to all 1s. Subsequently, the victim rows, now containing random data, are hammered. This is followed by reinitializing the attacker rows to all 0s and, once again, hammering the victim rows with the same random values. This cycle is repeated 200 times. For each cycle, we note down the bit flips on the victim rows. These bit flips are analyzed in a later stage of the attack.}

\hl{In the online attack scenario, the flip rates might be different than the offline profiling phase because of the presence of the victim process in the memory. To achieve more realistic flip rates, the offline profiling phase can be performed with an attacker's copy of the victim process located at the victim rows. 
 (Section~\ref{sec:memprof})}

\subsubsection{Amplifying Leakage} \label{sec:reclaiming}
\begin{figure}
    \centering
    \includegraphics[width=\columnwidth]{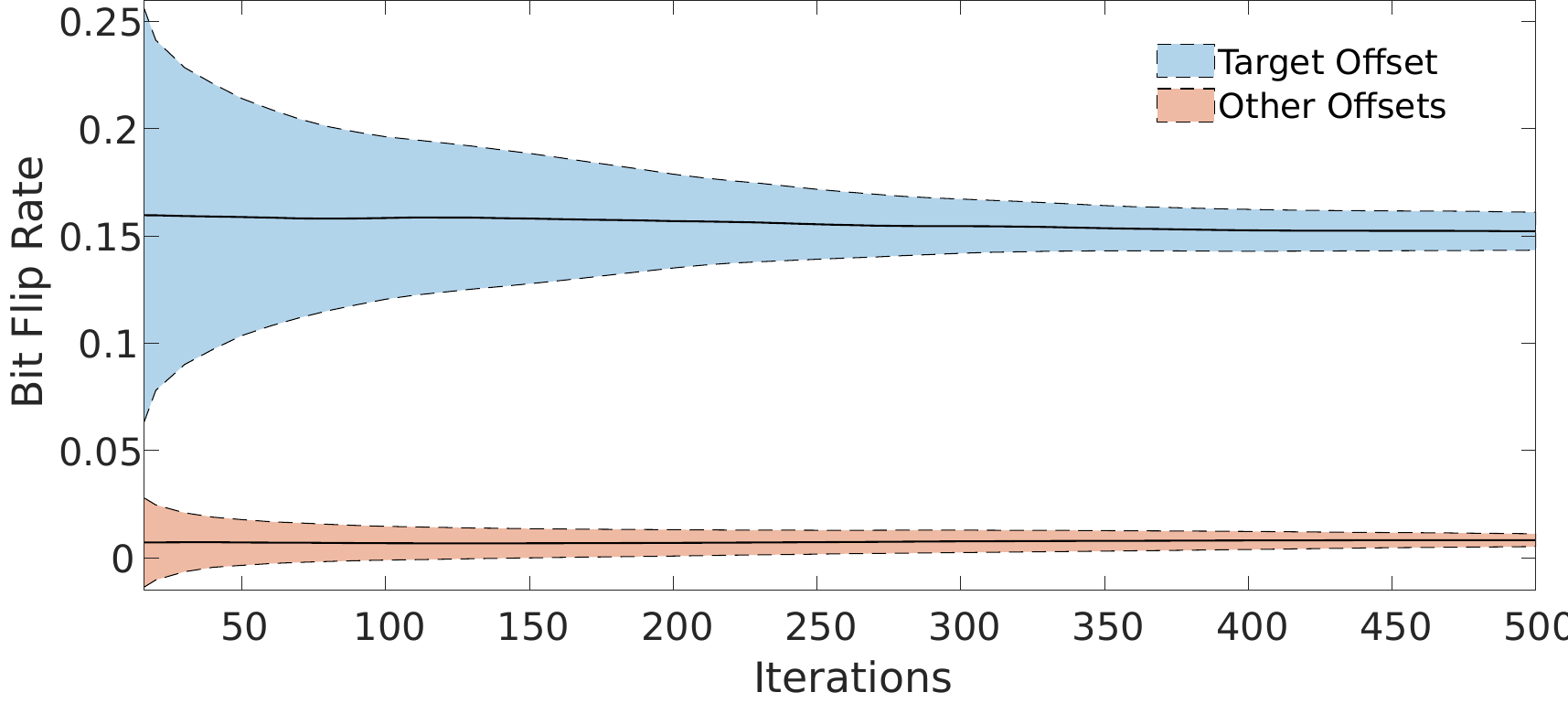}
    \caption{The variance and mean of bit flip rates for the target offset and other offsets in a single page for an increasing number of iterations. The separation between the target and other offsets allows us to distinguish the two based on the observed rate easily.}
    \label{fig:fliprate}
\end{figure}

The pages with page offsets that flip at every hammering attempt are rare. Therefore, a single Rowhammer session can only reveal the secret value with a certain probability. Moreover, the number of noisy bit flips within the page and other probabilistic factors, such as memory massaging, lower the probability of correctly predicting the secret bit. Hence, we need to amplify the probability by repeating the attack on the same secret value multiple times. Fig.~\ref{fig:fliprate} shows that there is a high variance in how many times a bit flip will be observed in a given Rowhammer iteration. The profiled page in this figure shows a flip characteristic where a flip in a given target offset is more likely to occur than any other offset.
After 200 iterations in the memory profiling, we observe that the variance in the flip rates of both the target and other offsets converges. Therefore, we use 200 iterations in the following experiments for the memory profiling and online attacks.

Once the attacker profiles the memory and finds relatively less noisy pages, these pages are unmapped, so the secret in the victim process is placed on the same flippy page. Yet, reclaiming the same physical page without any privilege is challenging due to virtual to physical address translations, which are opaque to users.

Here, we propose a novel method to reclaim the flippy pages without using Linux \texttt{pagemap}, which requires root privileges or huge pages that require special system configurations. 
\begin{itemize}[nosep,leftmargin=*]
    \item After an iteration of the online attack is completed, the victim program releases the flippy page. 
    \item Then, we allocate a large buffer of pages. As the page frame cache provides the recently used pages to the new allocations, the flippy page is highly likely to come back within the buffer.
    \item We cannot see the physical address of the flippy page to search within the buffer; therefore, we run another round of Rowhammer on the same attacker rows that we used for the online attack and search for a bit flip in the large buffer.
    \item Since we know the original flippy page offsets in the previously released page, we can then compare those flippy offset numbers with the actually flipped offset in the buffer. 
    \item If we cannot see a flip in the buffer in the first trial, we try hammering more. If it still fails, we increase the size of the buffer by mapping more pages and repeating the process until we can reclaim the victim page.
\end{itemize}
In our experiments on the DDR3 system, we observed that if the flippy page is in the allocated buffer, we can detect it by seeing a flip in the first few trials of Rowhammer, if not in the first trial.

\section{Experiment Results}

\paragraph{Experiment Setup}\label{para:setup}
The experiment is performed on an Ubuntu 20.04.6 LTS system with an Intel Core i5-8400 CPU @ 2.80 GHz and a Corsair DDR4 DRAM module with part number CMU64GX4M4C3200C16.

To execute our online attack, we employ offline memory profiling on the target system to pinpoint the most effective pages for the attack. 
Fig.~\ref{fig:flipvsbanks} shows the number of flips observed on different configurations. Increasing the number of banks increases the number of pages hammered and the number of flips observed. Moreover, the number of attackers significantly affects the number of observed flips on the same memory region.  

We profile the memory with different configurations on the number of attacker rows and the number of banks, i.e., \textbf{(R, B)}. We choose \textbf{(15, 7), (15, 5), (15, 4), (12, 7)} for the offline memory profiling phase. \textbf{(15, 7)} and \textbf{(15, 5)} configurations result in similar number of bit flips while \textbf{(15, 4)} configuration has less number of bit flips. We also use \textbf{(12, 7)} configuration to compare the result from a configuration that is comparably less as shown in Fig~\ref{fig:flipvsbanks}.



    \begin{figure*}
        \centering
        \begin{subfigure}[b]{0.475\textwidth}
            \centering
            \includegraphics[width=\textwidth]{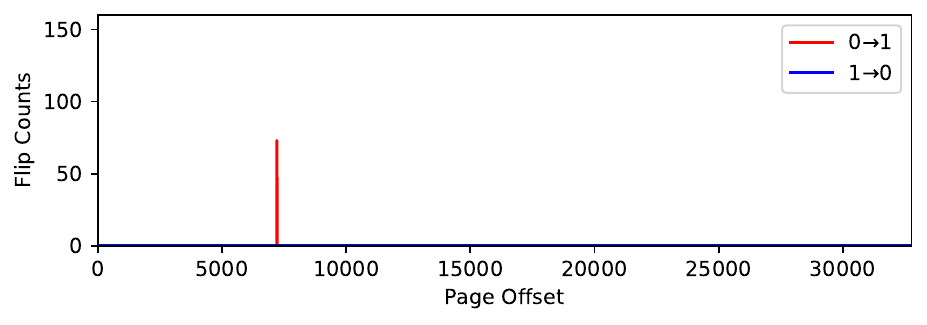}
            \caption[]%
            {{\small}}    
            \label{fig:a}
        \end{subfigure}
        \begin{subfigure}[b]{0.475\textwidth}  
            \centering 
            \includegraphics[width=\textwidth]{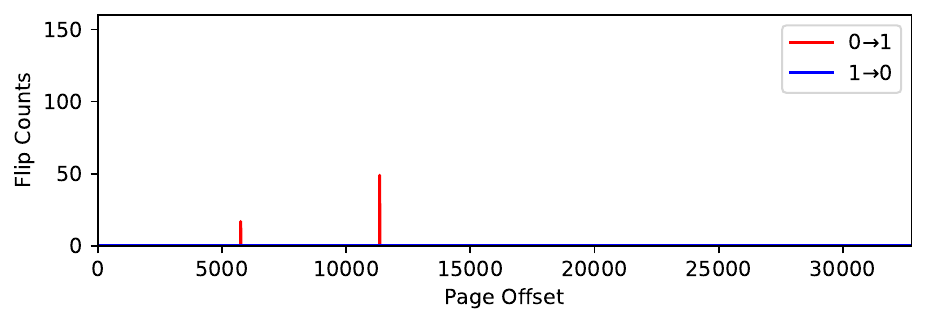}
            \caption[]%
            {{\small}}    
            \label{fig:b}
        \end{subfigure}
        \begin{subfigure}[b]{0.475\textwidth}   
            \centering 
            \includegraphics[width=\textwidth]{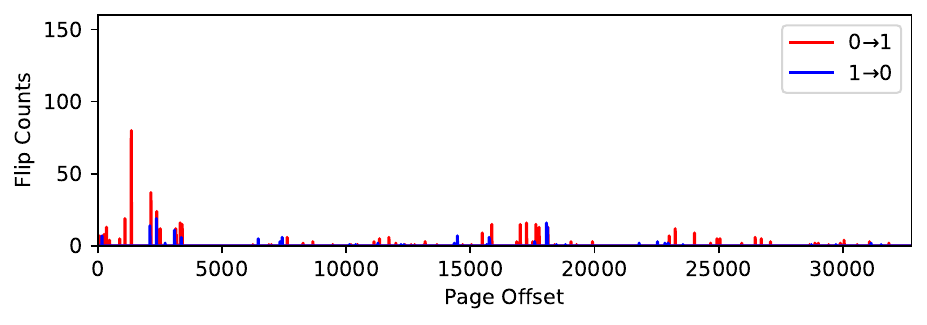}
            \caption[]%
            {{\small}}    
            \label{fig:c}
        \end{subfigure}
        \begin{subfigure}[b]{0.475\textwidth}   
            \centering 
            \includegraphics[width=\textwidth]{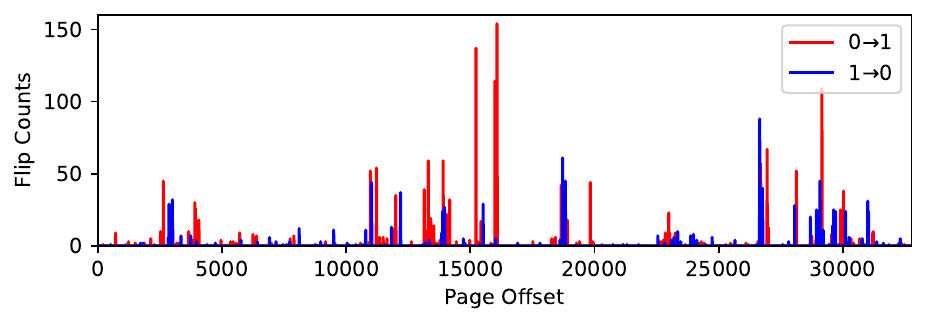}
            \caption[]%
            {{\small}}    
            \label{fig:d}
        \end{subfigure}
        \caption{Location and direction of bit flip on different pages. (a) and (b) are examples of \textit{reliable} pages, (c) and (d) are examples of \textit{unstable} and \textit{unusable} pages, respectively.}%
        \label{fig:diffpages}
    \end{figure*}

The offline memory profiling phase is performed as explained in Section~\ref{subsec:profiling} with chosen \textbf{(R, B)} configurations. Table~\ref{tab:flip_pages} shows the number of flippy pages under different profiling configurations. A page is referred to as a flippy page if at least one bit-flip is found during the offline memory profiling phase.

\hl{We observe that profiling the same physical address with the same hammering configuration results in the same bit-flip pattern. The number of bit flips might differ in different offline profiling phases, whereas the most flippy bit location stays the same. Furthermore, the hammering configuration affects the bit flip pattern on a physical address. The most flippy bit location changes for the same physical address with different hammering configurations. Thus, we can use the same physical address to probe different bits by using different hammering configurations. Some of the flippy pages in Table~\ref{tab:flip_pages} are the same physical addresses with different flip characteristics.}

\begin{table}
\footnotesize
        \centering
    \caption{Analysis of flippy pages in various hammering configurations that highlights how specific configurations impact the likelihood of inducing flippy pages.}

    \begin{tabular}{cccc}
    \toprule
             Configuration (R, B) & Profiled Area (MB) & \thead{\# Flippy \\{Pages}} & \thead{Flippy Page \\{/ MB}} \\
    \midrule  
             (15, 7) & 340.38 & 29,085 & 85.44 \\
             (15, 5) & 296.25 & 27,270 & 92.05 \\
             (15, 4) & 46.50 & 4,788 & 102.96 \\
             (12, 7) & 32.70 & 2,827 & 86.45 \\
    \bottomrule
        \end{tabular}
    \label{tab:flip_pages}
\end{table}

We profile more memory area with \textbf{(15, 7)} configuration than \textbf{(15, 5)} configuration. However, the number of flippy pages per MB is lower in \textbf{(15, 7)} configuration compared to \textbf{(15, 5)} configuration. Different parts of memory are allocated for offline memory profiling in different configurations. We allocate memory area less prone to bit flips in \textbf{(15, 7)} configuration. 

We profile 42\% more area with \textbf{(15, 4)} configuration than \textbf{(12, 7)}. We find 69\% more flippy pages in \textbf{(15, 4)} configuration. This also lays emphasis on the memory area allocated for the offline profiling phase.

\begin{table}
        \centering
    \caption{Distribution of bit flips across physical addresses, highlighting the most flippy bit location and the comparative number of flips on the identified flippy bit versus other bits.}
    \begin{tabular}{cccc}
    \toprule
             \thead{\# Physical \\{Address}} & \thead{Most Flippy \\{Bit Location}} & \thead{\# Flips on \\{Target Offset}} & \thead{\# Flips on \\{Other Offsets}} \\
    \midrule  
             2c0046000 & 2218 & 116  & 0  \\ 
             2bfa62000 & 7210 & 73  & 0 \\ 
             1ca224000 & 522 & 45  & 0 \\ 
             1e7598000 & 11347 & 49  & 17 \\ 
             135dd0000 & 7201 & 62   & 18 \\ 
             1cd7fc000 & 981 & 41   & 37 \\ 
             2c0ae1000 & 26034 & 42   & 28 \\ 
             2c1c8c000 & 1349 & 80   & 605 \\ 
             2c1d12000 & 9049 & 172   & 1153 \\ 
             2c1d10000 & 16076 & 154   & 2861 \\ 
    \bottomrule
        \end{tabular}
    \label{tab:page_flips}
\end{table}

Post-processing analysis is then applied to the collected bit flip data, producing Table~\ref{tab:page_flips} that illustrates the number of bit flips for random 10 flippy pages that were found in the profiling phase. Table~\ref{tab:page_flips} shows the most flippy bit location on each profiled DRAM page, where each DRAM page consists of 4KB, which is 32768 bits. The distribution of bit flips on DRAM pages varies, which is indicated by the number of total bit flips on the other bits. 

A detailed analysis of the distribution of flips on 4 different pages is shown in Fig.~\ref{fig:diffpages}. Some pages have distinctive bit flips on certain bit locations. It should be noted that we see the same bit flip direction on an offset while hammering it with 0 and 1. This would make it possible to probe the targeted bit regardless of the hammering values. Fig.~\ref{fig:diffpages} shows the total number of bit flips that occurred while hammering each page with 0 and 1. 
Fig.~\ref{fig:a} and~\ref{fig:b} show reliable bit flips on consistent bit offsets, while Fig.~\ref{fig:c} and~\ref{fig:d} have bit flips distributed over different offsets. Later, we split profiled pages into categories based on the number of bit flips. For our attack, we utilize pages with similar bit flip patterns as shown in Fig.~\ref{fig:a} and~\ref{fig:b}.

\paragraph{Experiment on Other DRAM modules}
We conduct a Rowhammer experiment on a different DRAM module for further analysis. We use a single bank multi-sided hammering configuration on a GSkill DDR4 DRAM module with part number F4-3600C16D-16GVKC.

Table~\ref{tab:flip_numbers_trr} shows the total number of bit flips found after hammering with different numbers of attacker rows. We observe that using more than 10 attacker rows decreases the number of flippy pages. 
The experiment result reasserts our statement about the impact of the hammering configuration on the number of flippy pages. Optimization of hammering configurations to perform offline memory profiling on other DRAMs requires the same steps explained in Section~\ref{sec:attack}.

\begin{table}
        \centering
    \caption{Analysis of flippy pages on GSkill DDR4 module with various hammering configurations that shows how specific configurations affect the number of flippy pages.}

    \begin{tabular}{cccc}
    \toprule
             \# Attacker Rows & Profiled Area (MB) & \thead{\# Flippy \\{Pages}} & \thead{Flippy Page \\{/ MB}} \\
    \midrule  
             13 & 1024 & 5,569 & 5.43 \\
             10 & 1024 & 17,746 & 17.33 \\
             9 & 1024 & 16,223 & 15.84 \\
    \bottomrule
        \end{tabular}
    \label{tab:flip_numbers_trr}
\end{table}

\section{Recovering wolfSSL TLS Server Key} \label{sec:wolfssl}

\begin{figure}[h]
    \centering
    \includegraphics[width=\linewidth]{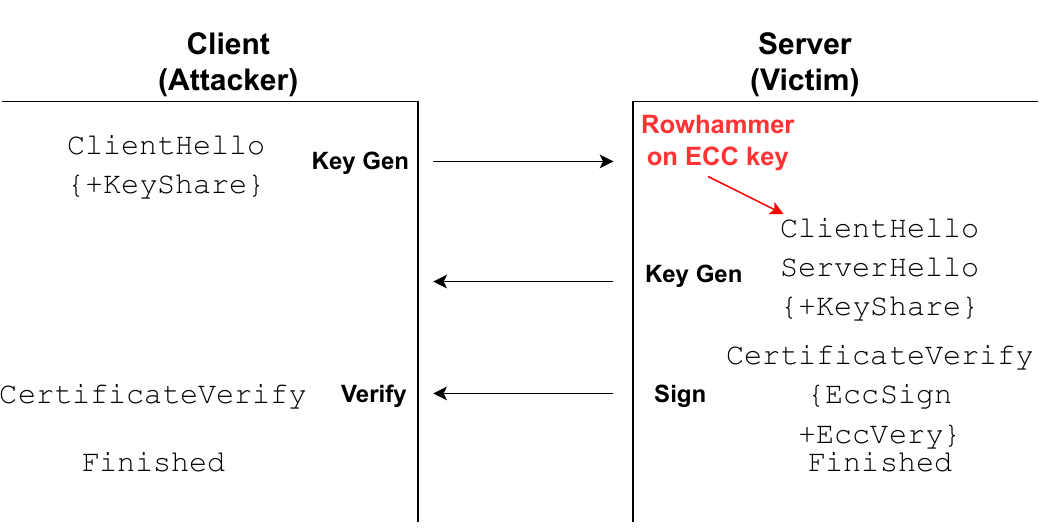}
    \caption{Rowhammer attack on TLS 1.3 handshake}
    \label{fig:tlshand}
\end{figure}

Our demonstration of the attack uses wolfSSL TLS 1.3 handshake protocol shown in Fig.~\ref{fig:tlshand}. It is important to note that in a practical attack scenario, offline memory profiling should cease once appropriate pages for the attack are found. Subsequently, the focus should shift to the online phase of the attack. For our online attack, we extensively profiled the memory setup to locate optimal pages for the attack.

Previously, wolfSSL addressed a Rowhammer attack that leveraged defective signatures to compromise the server's private key. The signature correction method applied to these signatures made it feasible to deduce the secret 256-bit ECDSA key~\cite{mus2023jolt}. As a countermeasure, wolfSSL added an additional verification step to the handshake protocol, involving a pre-send verification of signatures~\cite{wolfssl}. This step ensures that if a signature is found to be defective, the handshake, and consequently the SSL connection, does not occur. This step is shown as \textbf{EccVery} on the server side in Fig.~\ref{fig:tlshand}. This method is referred to as the \textit{verify-after-sign}. However, this fix inadvertently opens up ways for alternative attack strategies. Now, it is conceivable to infer the state of any targeted key bit by monitoring the failure rates of SSL connections during Rowhammer attacks. TLS 1.3 handshake protocol is summarized below.

\begin{itemize}[nosep,leftmargin=*]
    \item The client starts the handshake by sending \texttt{ClientHello{+KeyShare}} to the server.
    \item The server responds the client by sending \texttt{ServerHello{+KeyShare}} to the client. It also calculates the signature of a handshake and sends it to the client under \texttt{CertificateVerify} step. Optionally, it verifies the signature before sending it to the client. Thus, any defective signature is not sent to the client before the termination of handshake. Signature verification is optional and added as a countermeasure against a prior attack~\cite{mus2023jolt}.
    \item Once the client receives the response, it calculates its signature and verifies it. Then, the handshake is completed, and the secure connection between the client and the server is established.
\end{itemize}

If an attacker successfully injects fault into the ECDSA key, there are two possible ways to recover the secret bit.

\begin{itemize}[nosep,leftmargin=*]
    \item If the attacker gets the faulty signature from the server, a signature correction method can be used to recover a bit from the ECDSA key~\cite{mus2023jolt}.
    \item If the attacker does not get the faulty signature and observes that the handshake is terminated by the server, it is possible to recover a secret bit by repetitively sending handshake requests and injecting faults to the ECDSA key. Then, the attacker needs to observe the connection handshake termination rate to recover a bit from the ECDSA key.
\end{itemize}

\subsection{Memory Profiling Phase}\label{sec:memprof}

To provide a comprehensive summary of the offline memory profiling phase, we divide flippy pages found in Table~\ref{tab:flip_pages} into different categories based on the distribution of flips. Before categorizing the pages, we take sample pages from the flippy pages, which have bit flips distributed as shown in Fig.~\ref{fig:diffpages}. Then, we observe the handshake failures over different pages and categorize them to determine a criterion for pages that are useful for the online attack.

We define $\delta$ as the most number of bit flips observed on a single bit offset on a flippy page. We define $\sigma$ as the total number of bit flips observed on all bit offsets on a flippy page except the most flippy bit location. 

Some flippy pages have a very low number of bit flips, i.e., $\delta \leq 10$ and $\delta \geq \sigma$. On the other hand, some flippy pages have a higher number of bit flips where $\sigma \geq 80$. We observe that injecting a reproducible fault on the ECDSA key on these pages is not possible. We either see no bit-flips on the target offsets or lots of bit-flips on the other bit offsets. Thus, we refer to these pages as \textbf{unusable pages}.

We use flippy pages where $\delta \geq 10$ and $\delta \geq \sigma$ and perform an online attack. We observe that the handshake is terminated, and bit flips occur on the targeted bits on these pages. We categorize these flippy pages where $\delta \geq 10$ and $\delta \geq \sigma$ as \textbf{reliable pages}.

We use flippy pages where $\delta \geq 10$ and $\sigma \leq 80$ to perform the online attack. We observe that these pages can be used for the attack. However, they do not always result in consistent bit flips on the ECDSA key. Thus, it might give us a wrong conclusion of the real bit value. We categorize these pages as flippy pages where $\delta \geq 10$ and $\sigma \leq 80$ as \textbf{unstable pages}. 

Table~\ref{tab:profiled_pages} shows the profiled pages under different categories. There are more \textbf{reliable} and \textbf{unstable} pages in \textbf{(15, 7)} configuration compared to the \textbf{(15, 5)} configuration. This shows that \textbf{(15, 7)} configuration profiles pages that have consistent and reproducible bit-flips as shown in Fig.~\ref{fig:a} and~\ref{fig:b}.

\begin{table}
\footnotesize
        \centering
    \caption{Classification of memory pages into reliable, unstable, and unusable categories based on bit flip distribution across different hammering configurations.}
    \begin{tabular}{cccc}
    \toprule
             \thead{Configuration \\{(R, B)}} & Reliable Pages & Unstable Pages & Unusable Pages \\
    \midrule  
             (15, 7) & 685 & 2076 & 26324\\
             (15, 5) & 218 & 570 &  26482\\
             (15, 4) & 24 & 56 & 4708\\
             (12, 7) & 33 & 128 & 2666\\
    \bottomrule
        \end{tabular}
    \label{tab:profiled_pages}
\end{table}

\textbf{(15, 4)} configuration have less number of \textbf{reliable} and \textbf{unstable} pages compared to the \textbf{(12, 7)} configuration. \textbf{(15, 4)} configuration profiles more flippy pages. However, it either induces a very low number of bit-flips or a very high number of bit-flips on flippy pages, which are referred to as \textbf{unusable}. Therefore, we get more useful pages for the online attack from the \textbf{(12, 7)} configuration.

This analysis helps identify potential pages targeting specific bit offsets, which we repeatedly hammer in our online attack phase.

\paragraph{Extending the Offline Profiling for Higher Confidence}  \hl{The presence of the wolfSSL TLS server in the online attack causes some of the profiled pages to have lower flip rates than the offline profiling phase. This difference decreases the confidence in our probing process. 
Having low confidence in probed bits may still result in full key recovery on some schemes, such as on RSA with specialized algorithms~\cite{kwong2020rambleed}. Since other targets do not have specialized key recovery algorithms from noisy bits, we need to increase the confidence in the recovered bits.
For this reason, we extend the offline profiling phase.} 

\hl{We profile the \textbf{reliable} and \textbf{unstable} pages while an attacker's copy of the wolfSSL TLS server is located at them. The pages are loaded with an ECDSA key in which the probe offset is set to 0, and they are profiled as explained in Section~\ref{subsec:profiling}. Unlike the online attack phase, we do not need to connect to the server. 
After profiling with a 0 probe offset value, we restart the server on the same pages and load them with an ECDSA key in which the probe offset is set to 1. Then, we profile them to observe the bit flips on the probe offset. 
If the probed bit offset is still flippy, we use this page in the online attack phase.
}
\vspace{5mm}
\begin{table}[h!]
\centering

\caption{The effect of wolfSSL TLS server on the flip rates. If the probe offset is still flippy, the page is suitable.}
\resizebox{\columnwidth}{!}{%
\begin{tabular}{l|c|cc|c}

Physical Addr. & \thead{Offline\\Profiling} & \thead{ECDSA\\ Key Bit} & $\#$ of Bit Flips & Suitable?\\ \cline{1-5}
\multirow{ 2}{*}{2bf774000}  & 0 $\rightarrow$ 1: 17   & 0   & 0 $\rightarrow$ 1: 6 & \multirow{ 2}{*}{\cmark}    \\
  &   1 $\rightarrow$ 0: 0 & 1 & 1 $\rightarrow$ 0: 0     &  \\ \hline
\multirow{ 2}{*}{1d0220000}  & 0 $\rightarrow$ 1: 12  & 0 & 0 $\rightarrow$ 1: 0   &  \multirow{ 2}{*}{\xmark}  \\
 &1 $\rightarrow$ 0: 0  & 1 & 1 $\rightarrow$ 0: 0 &    \\ \hline
\end{tabular}%
}
\label{tab:extracted_pages}
\end{table}
\vspace{5mm}

\hl{Table~\ref{tab:extracted_pages} shows two pages to probe the same bit on the ECDSA key. The page at \texttt{2bf774000} has 17 instances of 0$\rightarrow$1 bit flips at the probe offset during profiling without the wolfSSL TLS server. We set the ECDSA key bit to 0 by loading the proper ECDSA key, and we profile the page. We observe 6 instances of 0$\rightarrow$1 bit flips, which is compatible with the profiling without the wolfSSL TLS server. Then, the ECDSA key bit is set to 1, and we profile the page again. We do not observe any bit flips this time, as expected. Since the probed bit offset is still flippy, this page is suitable for the online attack phase.}

\hl{The page at \texttt{1d0220000} has 12 instances of 0 $\rightarrow$ 1 bit flips on the probe offset during profiling without the wolfSSL TLS server. We repeat the same steps as the previous page.
Since the probed bit offset is not flippy, this page is not suitable for the online attack phase. All \textbf{reliable} and \textbf{unstable} pages are profiled in this manner to find the suitable pages needed to perform the online attack.}

\subsection{Online Attack Phase}\label{sec:online_attack}

\paragraph{Allocating the Useful Pages} Once we identify pages that are useful to our attack, the next step involves allocating these pages for use in our online attack. For experimental purposes, during the profiling stage, we record the physical addresses of the victim pages and their corresponding attacker pages. We utilize the \texttt{mmap} system call to reallocate these pages, allocating various memory segments. We then verify the physical addresses of these allocated memory pages. Pages with matching memory addresses are retained for use in the attack, while those that don't match the required physical address values are released. 

It is important to note that this memory allocation step would not be necessary in a full-scale, end-to-end attack. Should an appropriate page be identified during the profiling stage, the next action would be to proceed with the online phase of the attack immediately.

\paragraph{Mapping Private Key to Victim Page} To successfully flip bits in the ECDSA key, the memory arrangement must be modified, ensuring the key is positioned in one of the previously identified vulnerable rows. This is achieved by initially unmapping the areas prone to flips, after which the client either generates or loads the private key. Leveraging the characteristics of the Linux Buddy Allocator~\cite{bovet2005understanding}, which allocates newly freed physical pages from the page frame cache in a first-in-last-out manner, the private key gets mapped onto the vulnerable row.

Once the server initiates the TLS 1.3 handshake process, the ECDSA key is loaded from the key certificate to the server memory. In our attack scenario, we impersonate a client to establish a connection with the server, which is in a state of awaiting connection requests. A crucial aspect of the attack is the timing: it is imperative not to hammer the victim page until the key has been allocated to it. To address this timing or synchronization challenge, the client (in this case, the attacker) should delay the connection request, allowing sufficient time for the key to be generated and placed on the victim page. As an attacker, we have control over the timing of the connection request, enabling us to synchronize our actions with the server's key generation process.

Once we achieve the necessary synchronization, we hammer the attacker pages allocated prior to initiating the attack. Subsequently, we monitor the status of the TLS 1.3 handshake to determine whether a fault has occurred in the key. However, it is important to note that the connection status does not always provide direct insight into the state of the secret bit. There are two scenarios to consider:
\begin{itemize}[nosep]
    \item First, the secret key might not be located on our targeted victim page. In such instances, our bit flipping may impact other variables within the wolfSSL framework rather than the intended key. 
    \item Second, there's a possibility that the bits being flipped belong to memory allocations of other processes, not just wolfSSL. In such cases, a failure in the connection does not necessarily imply a compromise of the secret bit. 
\end{itemize}
This underscores the complexity of the attack and the need for careful analysis of the results to interpret the impact of our actions accurately.

\paragraph{ASLR Effect on the Key Allocation} Address Space Layout Randomization (ASLR) randomizes the victim memory space as a caution against buffer overflow attacks. If it is disabled by the system, we observe that the physical address of the page for the server key stays the same. This would increase the flips observed on the victim's system since we always target the same victim row. If ASLR is enabled, our chance of placing the server key into the victim page will decrease. 
However, it will only increase the time of our attack since it requires more iterations to eliminate the noise.

\begin{figure}
    \centering
    \includegraphics[width=0.8\linewidth]{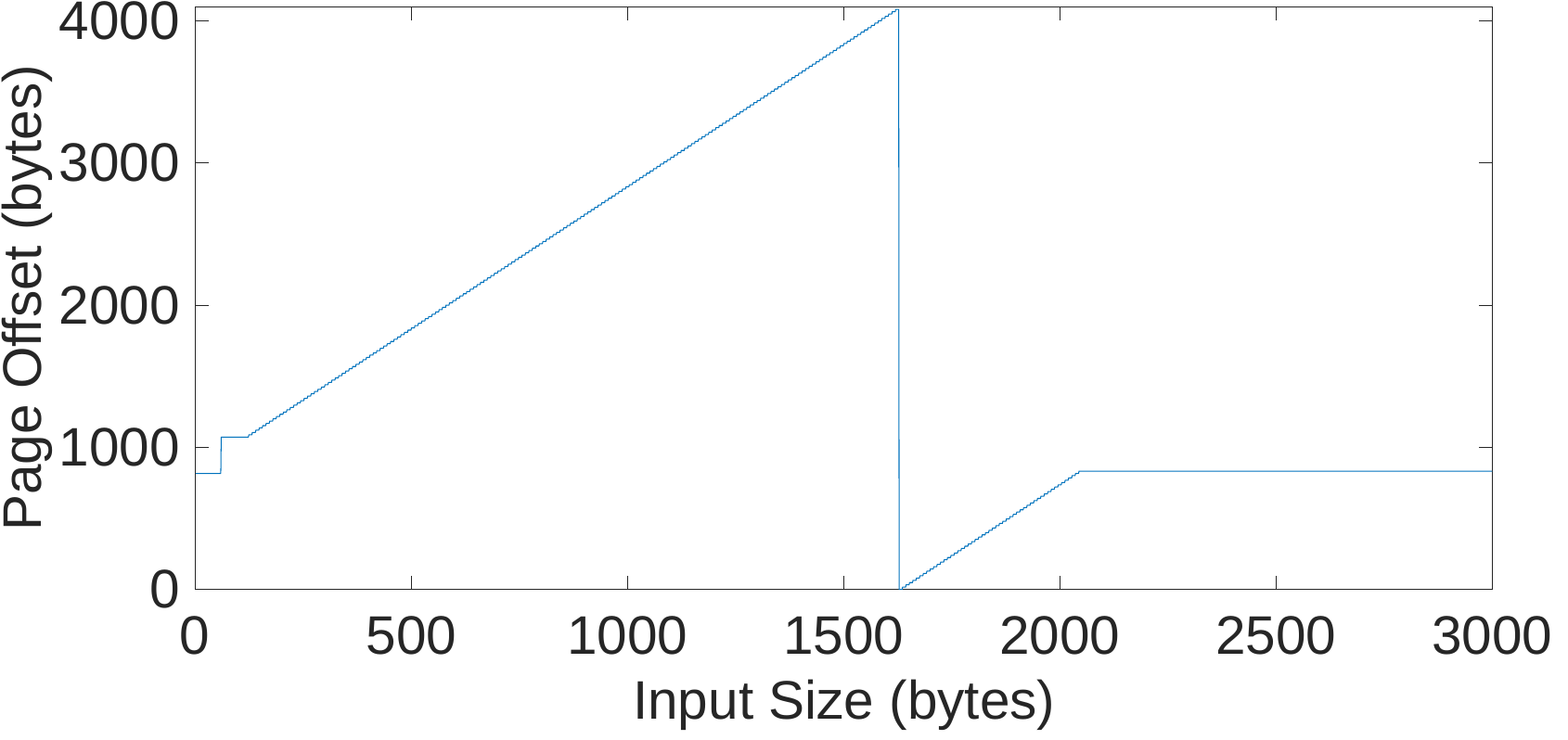}
    \caption{Shifting page offset of the secret using the attacker-controlled input size}
    \label{fig:malloc_shift}
\end{figure}

\paragraph{Shifting the Page Offset of Key} Our attack depends on the flippy bit offset on a page found in the offline memory profiling phase. Thus, it is vital to match the flippy bit offset with the bit offset of the server key. If any mismatch occurs, we will target a bit that is different than intended or belongs to the other variables in the process. We observe that the page offset of the server key always stays the same since ASLR does not randomize the page offset. In this case, we would limit our attack to the profiled pages with bit-flips only on a constant 256 bit-space over a 4KB page. 
To show the viable attack scenario, we utilize an input-dependent \texttt{malloc} size on a variable from the victim's side. This changes the page offset of the server key and enables us to place it into the desired offset on the victim page.

In case there is no attacker-controlled \texttt{malloc} before the secret, extending the offline memory profiling phase is still possible to find flippy pages for each bit offset. This would only prolong the offline attack time. 

\begin{lstlisting}[ backgroundcolor = \color{white},caption={An example of dynamic memory allocation for server TLS 1.3 application from wolfSSL 5.6.3},label={tlscode},frame=single,linewidth=0.981\columnwidth,xleftmargin=0.35cm]
int main(int argc, char** argv)
{
  int ret = 0;
#ifdef WOLFSSL_TLS13
  struct sockaddr_in servAddr;
  struct sockaddr_in clientAddr;
  socklen_t          size = sizeof(clientAddr);
  char               buff[256];
  size_t             len;
  int                size = atoi(argv[2]);
  char *reply = (char*) malloc(sizeof(char) *size);
  int                on;
  /* declare wolfSSL objects */
  WOLFSSL_CTX* ctx = NULL;
  WOLFSSL*     ssl = NULL;
  ...
  return ret;
}
\end{lstlisting}
\vspace{7mm}

On code snippet List.~\ref{tlscode}, \texttt{reply} variable is allocated by using \texttt{malloc}. This would affect the bit location of the ECDSA key, which is stored in \texttt{ctx} variable. As the size of \texttt{malloc} increases by changing \textbf{size} variable, the location of the ECDSA key within the DRAM page shifts toward the next DRAM page. We illustrate in Fig.~\ref{fig:malloc_shift} that an attacker who is in control of the size of a \texttt{malloc} can indeed shift the secret variable around the page boundaries, spanning all possible page offsets.
This would allow us to utilize every \textbf{reliable} and \textbf{unstable} pages found in Section~\ref{sec:memprof}.

\hl{The \texttt{malloc} system call allocated memory that is 128-bit aligned. It allows us to shift \texttt{ctx} variable by 128-bit on the same page. This gives us the opportunity to probe congruent bits in the ECDSA key i.e. 0$^{th}$ - 128$^{th}$, 1$^{st}$ - 129$^{th}$ bits, and so on. Hence, we only need 128 pages to probe the entire 256-bit ECDSA key.}

\paragraph{Performing the Online Attack} \hl{Each page used for the online attack is hammered with the attacker pages set to 1s, and any connection failure is observed by connecting to the server as a client. This cycle is repeated 200 times. Then, the page is hammered with 0s to see if any connection failure happens. We also repeat this cycle 200 times. Then, we shift the ECDSA key 128-bit and repeat the attack cycle to probe the congruent bit on the key. Each page is used 800 times for the online attack to probe two bits on the ECDSA key.}

\subsection{Online Attack Results}

\begin{table}
        \centering
    \caption{Summary of online attack outcomes on a fixed ECDSA key, showing the relationship between offline profiling, connection failures, and bit value predictions versus actual values.}
    \resizebox{\columnwidth}{!}{%
    \begin{tabular}{c|ccccc}
    \toprule
             Pages & \thead{Physical\\{Address}}&  \thead{Offline\\{Profiling}} & \thead{Number of \\{Connection Failures}} & \thead{Probed\\{Bit Value}} & \thead{Real Bit\\{Value}} \\
    \midrule           
            \multirow{6}{*}{\rotatebox[origin=c]{90}{Reliable}}             
             & \multirow{2}{*}{2bf774000} & \multirow{2}{*}{0 $\rightarrow$ 1 = 17} & 6 & 0 & 0  \\
             &  &  & 0 & 1 & 1  \\ \cline{2-6}
             & \multirow{2}{*}{1cc3fc000} &  \multirow{2}{*}{1 $\rightarrow$ 0 = 14} & 0 & 0 & 0  \\
             &  &  & 6 & 1 & 1  \\ \cline{2-6}
             & \multirow{2}{*}{2c01db000} & \multirow{2}{*}{1 $\rightarrow$ 0 = 13} & 0 & 0 & 0  \\
             &  &  & 6 & 1 & 1  \\             
    \midrule
            \multirow{4}{*}{\rotatebox[origin=c]{90}{Unstable}}
            & \multirow{2}{*}{1b1cc5000} & \multirow{2}{*}{0 $\rightarrow$ 1 = 20}  & 4 & 0 & 0 \\
             &  &   & 0 & 1 & 1 \\ \cline{2-6}             
             & \multirow{2}{*}{2c06ad000} & \multirow{2}{*}{0 $\rightarrow$ 1 = 10} & 8 & 0 & 0  \\
             &  &  & 0 & 1 & 1  \\              
        \bottomrule
        \end{tabular}%
        }
    \label{tab:online_attack2}
\end{table}

\hl{We perform the online attack as explained in Section~\ref{sec:online_attack}. Table~\ref{tab:online_attack2} shows the online attack results on several examples of \textbf{reliable $\And$ unstable} pages.}

\hl{For example, the page located at \texttt{2bf774000} has 17 instances of 0$\rightarrow$1 bit flips during the offline profiling. We perform the online attack, and we observe 6 connection failures and probe the bit value as 0.
Then, we perform the online attack on the \textit{shifted} ECDSA key and do not observe any connection failures. Thus, we probe the bit value as 1 since we expect to see instances of 0$\rightarrow$1 bit flips that hinder the connection.  }

\hl{The page located at \texttt{1b1cc5000} has 20 instances of 0$\rightarrow$1 bit flips during the offline profiling. We perform the online attack and observe 4 connection failures. Thus, we probe the bit value as 1 since we expect to see 0$\rightarrow$1 bit flips that prevent the connection. On the \textit{shifted} ECDSA key, we do not observe any connection failures and probe the bit value as 0. After the attacks, we check the real bit values and validate our probing results.}

\hl{Overall, the attack is performed on a 256-bit ECDSA key, and we achieve an average recovery rate of 22 bits/hour with a 100\% success rate.}

\section{Comparison to Related Work}
\paragraph{Rowhammer-based KEM Attacks}
The works in \cite{FrodoKemattack} and \cite{LWERowhammer} both target PQ KEM schemes using Rowhammer fault injection. The former uses Rowhammer to increase the error rate during encryption in FrodoKEM, which results in higher decryption failures, which are then mathematically analyzed to deduce key bits. Similarly, ~\cite{LWERowhammer} achieves key recovery from LWE-based KEM schemes using Rowhammer by running the decapsulation procedure or the plaintext checking oracle multiple times with different ciphertexts while introducing faults that reveal parts of the key. Both techniques require extensive analysis. In contrast, \attack\ can directly fault and probe bits without any extensive analysis given an observation channel. 

\paragraph{Signature Correction Attacks} SCAs such as LUOV\cite{quantumhammer}, Dilithium~\cite{Islam2022SignatureCA} and EC/DSA~\cite{mus2023jolt}, on the other hand, recover key bits by correcting faulty signatures obtained by Rowhammer fault injection. Our attack does not require access to faulty signatures/ciphertext or any complex mathematical postprocessing to recover the secret key bit. For instance, for signature correction to work, the faulty bit needs to be mathematically \textit{traceable} to the faulty signature, allowing signature correction. Instead, the presented \attack attack can directly recover bit values either by observing signing rates (if fault checking is implemented) or by recovering error-dependent change in the protocol's behavior.

\paragraph{RAMBleed}
Kwong \etal~\cite{kwong2020rambleed} proposed RAMBleed to extract bits using Rowhammer. Our attack is fundamentally different in terms of how the secret bit is deduced. In RAMBleed, the attacker exploits the data dependency of the aggressor rows. \attack exploits the observable effect of a bit flip on the secret value.
Moreover, RAMBleed has three fundamental limitations that \attack does not have.

First, in RAMBleed, the attacker has to co-locate two copies of the victim in the same rows and force them to hammer the attacker's row. In our attack, however, the victim can exist in any row as long as the attacker rows cause a bit-flip on the victim row.
\vspace{5mm}
\begin{figure}[h]
    \centering
    \includegraphics[width=0.8\linewidth]{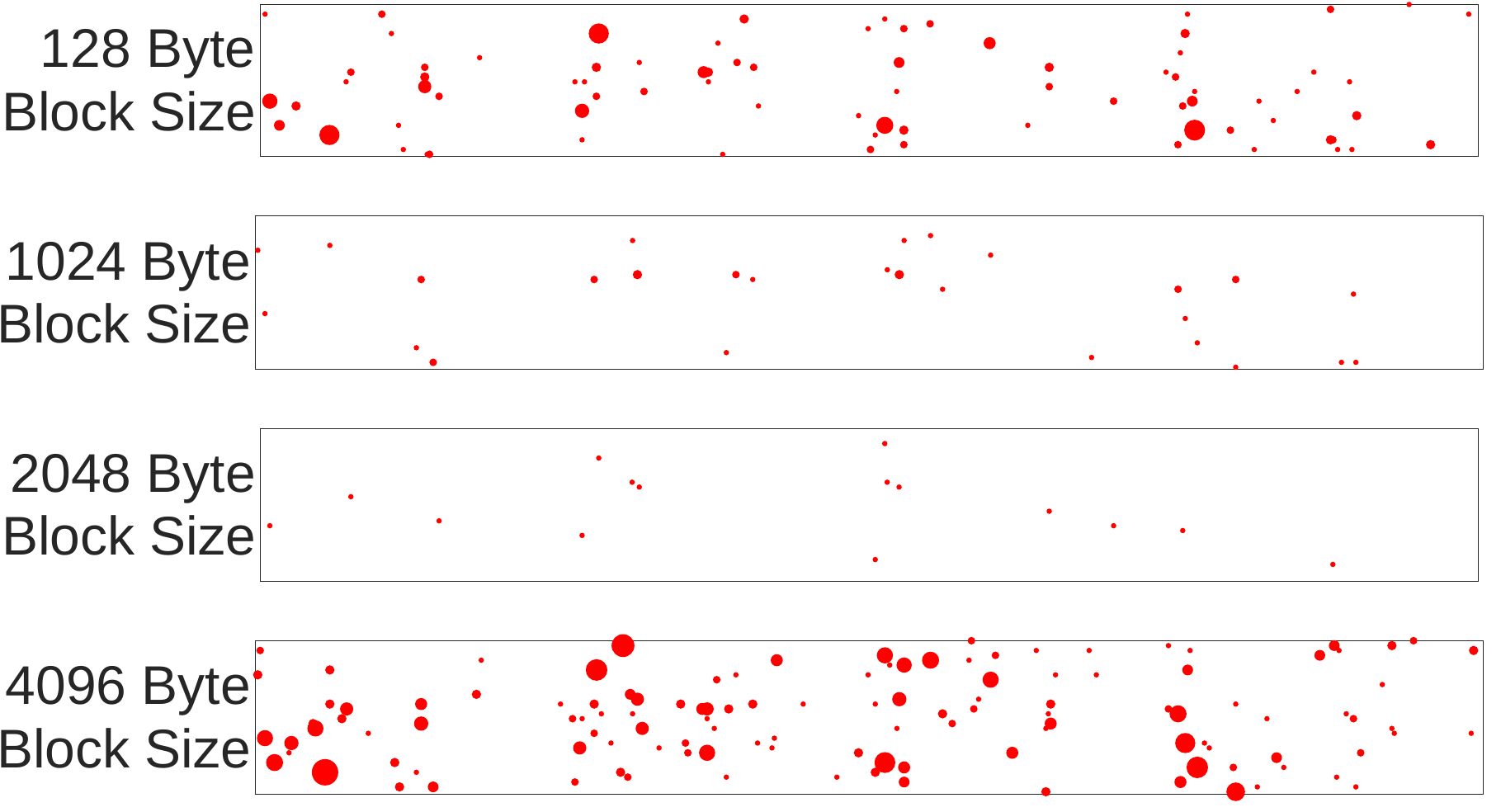}
    \caption{Flip characteristics of a single page with different attacker row data patterns. Larger dots indicate more repeatable bit flips in a certain data pattern in DDR4.}
    \label{fig:blocksize_scatter}
\end{figure}
\vspace{7mm}
\hl{Second, RAMBleed can only deduce the bits with a certain probability. Probabilistic information is enough to extract a full RSA key using a variant of Henninger-Shacham algorithm~\cite{heninger2009reconstructing, paterson2012coding}, yet it does not work on other schemes such as ECDSA. Since \attack can deduce bits with 100\% accuracy, it can be used as a generic attack on any secret that can be flipped using Rowhammer and reveals fault information.}

Finally, 
the RAMBleed attack was based on the assumption of data dependency of bit flips on DDR3 memory chips. 
It assumes that a bit flip strongly depends on the bit value of the same offset in the immediate neighbors. However, in our experiments on DDR4, we have observed that this is not the case. 
We tried different aggressor row patterns to show the effect of diagonal memory cells on a flippy page offset. We set the value of the hammering rows as $111..1000..0111..1$ where we call running 1s or 0s a \textit{block}. 
Fig.~\ref{fig:blocksize_scatter} demonstrates how a single page shows different flip characteristics when we change the size of the block.
We observed that different values on the attacker rows change these flip characteristics of the victim row entirely. Therefore, without knowing the actual value of the whole victim row, it would not be possible to comment on the flips seen on the attacker-controlled "sample page". Knowing every bit of the secret value prior to the attack would be an unrealistic assumption.
In conclusion, a profiling-based attack like RAMBleed would not work on DDR4 systems.

\section{Applying \attack to Other Libraries}
We have reviewed TLS applications from \texttt{OpenSSL}, \texttt{LibreSSL}, and \texttt{Amazon s2n} libraries to check the feasibility of \attack. We have performed an end-to-end attack on \texttt{WolfSSL}, including offline memory profiling and mapping the private key to the victim page. Since we would use the same DRAM modules for the other libraries, we expect to perform \attack and recover the private key. Therefore, in a simulated fault attack scenario, we only check how the private key is stored in these libraries.

\paragraph{OpenSSL 3.3.0}
\texttt{OpenSSL}, a widely-used open-source toolkit for TLS protocols, provides essential security features for TLS 1.3 applications. We verified that \attack is applicable to \texttt{OpenSSL 3.3.0} by using the handshake protocol in Fig.~\ref{fig:tlshand}. \texttt{OpenSSL} server reads the server key and waits for a handshake request from a client. Meanwhile, the server key is stored unmasked and vulnerable to Rowhammer attack. \attack can trace the private key bits and recover the server key.

The \texttt{OpenSSL} team verified the vulnerability reported in~\cite{mus2023jolt}, but they have not released patches to fix it, stating Rowhammer is not in their treat model. 

\paragraph{LibreSSL 3.9.2}
\texttt{LibreSSL}, a fork of the OpenSSL project, aims to provide a more secure and modern implementation of TLS 1.3 protocol. They verified the vulnerability reported in~\cite{mus2023jolt} and fixed it by performing~\textit{verify-after-sign} method in the later patches. However, we can still perform \attack on \texttt{LibreSSL} since they store the server key unmasked before getting a handshake request from a client, as shown in Fig.~\ref{fig:tlshand}.

\paragraph{Amazon s2n-TLS 1.4.15}
\texttt{Amazon s2n-TLS}, a lightweight and efficient implementation of TLS protocol, is designed to provide secure communications for applications and services. \texttt{Amazon s2n} confirmed the vulnerability reported in~\cite{mus2023jolt} and implemented~\textit{verify-after-sign} method as~\texttt{WolfSSL} and~\texttt{LibreSSL} adopted for their libraries. We can perform \attack on \texttt{Amazon s2n} that utilizes \texttt{OpenSSL} TLS 1.3 application. 
We verified that \texttt{Amazon s2n} stores the private key unmasked, and it can be recovered by \attack even under~\textit{verify-after-sign} fix is enabled.

\section{Responsible Disclosure}
We disclosed our findings to \texttt{wolfSSL} on April 4, 2024. 
\texttt{wolfSSL} team acknowledged the vulnerability. 
A comprehensive patch was released,
utilizing \textit{blinded ECDSA} with the masked private key to minimize key exposure. We tested \attack on the patched libraries and verified that the patch is effective in mitigating our attack.

\section{Countermeasures}
In this section, we discuss the viability of several countermeasures against \attack that can be implemented on both software and hardware.

\paragraph{Mitigating Faults}
In \attack, we rely on bit corruption using Rowhammer. Therefore, any defense mechanism that prevents targeted Rowhammer attacks also prevents \attack.

Reducing the DRAM refresh interval constrains the attack's potential by limiting the available time for inducing bit flips. However, this measure leads to increased power consumption and decreased system performance since memory cannot be accessed during the refresh operation, thereby making it an impractical solution for widespread adoption.

Researchers have proposed alternative row refresh techniques that aim to balance effectiveness against performance and energy penalties. One notable example is the Hidden Row Activation (HiRA)~\cite{HiRA2022}, which introduces a method for parallelizing row refreshes. HiRA capitalizes on the architectural feature that different rows within the same bank can be connected to distinct charge restoration circuits, enabling simultaneous refresh operations. This method not only diminishes the window available for Rowhammer attacks by embedding refresh operations into the normal row access cycles but also mitigates the associated latency issues. Although HiRA offers a promising approach to mitigating Rowhammer attacks, its integration into actual products is still forthcoming.

Further advancements are represented by the work of Wang \etal \cite{wang2021discreetpara}, which advances the Probabilistic Adjacent Row Activation (PARA)~\cite{kim2014flipping} defense mechanism through the development of Discreet-PARA. This approach integrates disturbance tracking with a dedicated cache for recently accessed rows (PARA-cache), facilitating precise management of refresh operations for rows at risk of Rowhammer attacks. Discreet-PARA has successfully reduced the performance overhead typically associated with such mitigation techniques by refining the process of monitoring row accesses and refreshes. Since PARA requires changes in the memory controllers or DRAM chips, it is not possible to implement them on the current systems~\cite{mutlu2017rowhammer}.

Although  Error Correcting Code (ECC) can correct single-bit flips, Rowhammer attacks capable of generating multiple-bit flips can circumvent ECC protection, as demonstrated in \cite{cojocar2019ecc}. This limitation underscores the necessity for Rowhammer-specific countermeasures even in systems equipped with ECC.

\paragraph{Mitigating Probes} For extracting the secret bit value, we rely on a feedback mechanism that the adversary observes. Eliminating or blinding this channel would mitigate \attack as well.

In server/client scenarios where a malicious client collects faulty signatures, verifying the signature before sending it to the client would prevent the client from deducing secret bits \textit{only by using the signature verification.} Yet, as we explained in the previous sections, this alone is not a sufficient mitigation.
Even if the server implements verify-after-sign as a security measure, how it handles an unverified signature scenario is also important for \attack. Libraries need to prevent error codes from revealing information to the clients if there is a fault in the secret values. This requires a thorough review of the error codes.

Finally, not releasing any error code to the client and correcting the fault on the fly can be the last resort. However, an error handling mechanism needs to be implemented to be constant-time (or, more precisely, fault-independent). For example, repeating the signing operation when the verify-after-sign fails can potentially cause a visible delay in the connection, and the malicious client can deduce that the fault went through successfully. To prevent that, \hl{two copies of the private key can be stored in separate locations in the memory and} two signatures can be generated simultaneously, and whichever is verified can be sent to the client. This would cause computational overhead, but if implemented constant-time, the client would not see the effect of the fault. \hl{However, there is a small chance that both copies of the key can be flipped on the same bit offset, which may reveal the fault information.}
Furthermore, an unmasked secret key may be vulnerable to probing tools. Libraries need to store sensitive information in masked form to prevent real secret bits from being recovered.
Specifically, ECDSA implementations are vulnerable to \attack unless secret key blinding is implemented to mitigate our attack.

\section{Exploiting ASLR as an Attack Vector}
When the number of reliable pages that can generate reproducible bit flips with low noise is limited, we need to shift the secret value within a page to reuse the same pages to probe multiple bits. In Section~\ref{sec:online_attack}, we explained that this can be done using an attacker-controlled \texttt{malloc} size.

Previous works~\cite{adiletta2023mayhem, tobah2022spechammer} demonstrated flipping variables stored in stack memory.
In case the secret value is stored in stack memory instead of heap memory, changing the \texttt{malloc} size will not affect the page offset of the secret. Yet, ASLR implemented in the Linux kernel randomizes page offset by default. Here, we propose to exploit address randomization that is implemented to improve performance by \textit{``avoiding L1 evictions by the processes running on the same package''}~\cite{vandeVen2005RandomizeStackPointer,Molnar2006EnhancingKernel} to enable \attack against the secret variables that are stored in stack memory. This is critical when the number of reliable pages is not enough to cover the secret size.

\begin{figure}
    \centering
    \includegraphics[width=\linewidth]{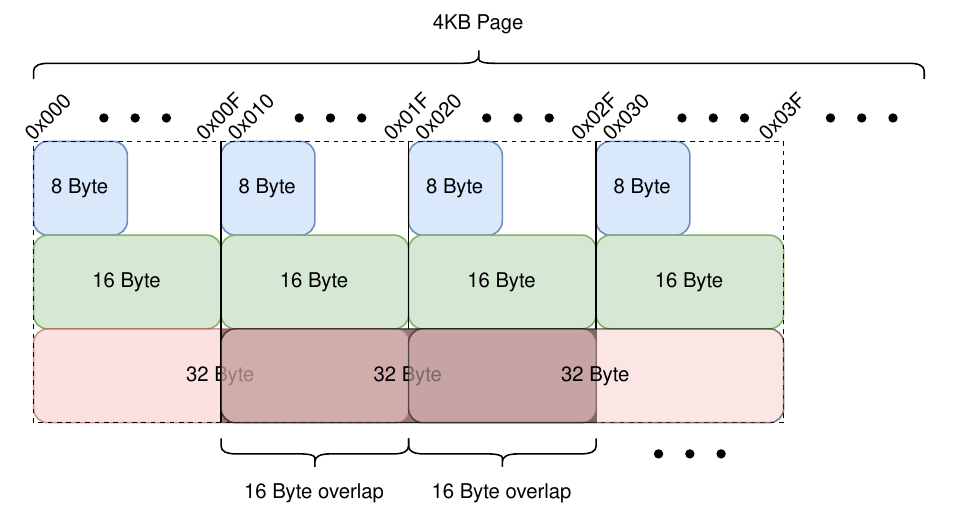}
    \caption{ASLR Shift overview}
    \label{fig:aslr_shift}
\end{figure}

Fig.~\ref{fig:aslr_shift} shows an overview of the ASLR shift idea. Note that regardless of the size of the variable, the last 4 bits of the address are not affected by the offset randomization. Yet, the remaining 8 bits of the page offset are randomized.
Let's say a variable has a page offset \texttt{addr} in one run and \texttt{addr}$+$\texttt{0x10} in the second run.
Since randomization does not affect the last four bits of the address if the variable's size is less than 16 bytes, all bits in the variable will always be mapped into unique locations on the physical page. 
Yet, if the variable is larger than 16 bytes, \texttt{addr} and \texttt{addr}$+$\texttt{0x10} will have overlapping bits for different runs.

Assuming that an unprivileged adversary has no prior information about the randomized page offset of a variable in a given run, using our \attack, they can deduce where this flip occurred based on the unique mapping of every bit for variables with a size smaller than 16 bytes. Similar to the technique explained in Section~\ref{sec:online_attack}, since the flip characteristic of the profiled page is known, an error in the program will indicate the original value of the corrupted bit.

For variables with a size larger than 16 bytes, the value of the secret bit can be found the same way. However, the overlapping parts decrease the probability of deducing the location of the bit within the variable. For example, for a 32-byte variable, a detected bit flip at offset \texttt{0x020} may be located either at the beginning of the first or second half of the variable, resulting in two different options.

To generalize, in a variable with size $n$ bytes, an adversary can tell the location of each corrupted bit on the variable with a $16/n$ probability where $n\geq16$ and $n \equiv 0 \pmod{16}$.

\vspace{7mm}

\section{Conclusion}
In this study, we presented \attack, a novel methodology that uses the Rowhammer to deduce secret information exploiting the observable changes in a system's behavior following fault injections. Through rigorous experimentation, including a targeted attack on the ECDSA signature scheme in wolfSSL TLS handshake, we have demonstrated the capability of \attack to bypass established fault-check mechanisms.

The introduction of \attack underlines the necessity for a refined defensive strategy against Rowhammer and similar hardware-level threats.
Future countermeasures must address the attack vectors that exploit operational behaviors as side channels.

\section*{Acknowledgment} 
This work was supported by the National Science Foundation grant CNS-2026913 and in part by a grant from the Qatar National Research Fund.


\bibliographystyle{plain}
\bibliography{references}

\begin{thebibliography}{10}

\bibitem{adiletta2023mayhem}
Andrew~J. Adiletta, M.~Caner Tol, Yarkın Doröz, and Berk Sunar.
\newblock Mayhem: Targeted corruption of register and stack variables.
\newblock In {\em Proceedings of the 2024 ACM Asia Conference on Computer and Communications Security}, 2023.

\bibitem{aweke2016anvil}
Zelalem~Birhanu Aweke, Salessawi~Ferede Yitbarek, Rui Qiao, Reetuparna Das, Matthew Hicks, Yossi Oren, and Todd Austin.
\newblock {\MakeUppercase{ANVIL}}: Software-based protection against next-generation rowhammer attacks.
\newblock {\em ACM SIGPLAN Notices}, 51(4):743--755, 2016.

\bibitem{biham1997differential}
Eli Biham and Adi Shamir.
\newblock Differential fault analysis of secret key cryptosystems.
\newblock In {\em Advances in Cryptology—CRYPTO'97: 17th Annual International Cryptology Conference Santa Barbara, California, USA August 17--21, 1997 Proceedings 17}, pages 513--525. Springer, 1997.

\bibitem{bovet2005understanding}
Daniel~P Bovet and Marco Cesati.
\newblock {\em Understanding the Linux Kernel: from I/O ports to process management}.
\newblock " O'Reilly Media, Inc.", 2005.

\bibitem{brasser2017cant}
Ferdinand Brasser, Lucas Davi, David Gens, Christopher Liebchen, and Ahmad-Reza Sadeghi.
\newblock {CAn{\textquoteright}t} touch this: Software-only mitigation against rowhammer attacks targeting kernel memory.
\newblock In {\em 26th USENIX Security Symposium (USENIX Security 17)}, pages 117--130, Vancouver, BC, August 2017. USENIX Association.

\bibitem{canella2019fallout}
Claudio Canella, Daniel Genkin, Lukas Giner, Daniel Gruss, Moritz Lipp, Marina Minkin, Daniel Moghimi, Frank Piessens, Michael Schwarz, Berk Sunar, Jo~Van~Bulck, and Yuval Yarom.
\newblock Fallout: Leaking data on meltdown-resistant cpus.
\newblock In {\em Proceedings of the 2019 ACM SIGSAC Conference on Computer and Communications Security}, CCS '19, page 769–784, New York, NY, USA, 2019. Association for Computing Machinery.

\bibitem{chakraborty2020explframe}
Anirban Chakraborty, Sarani Bhattacharya, Sayandeep Saha, and Debdeep Mukhopadhyay.
\newblock Explframe: exploiting page frame cache for fault analysis of block ciphers.
\newblock In {\em 2020 Design, Automation \& Test in Europe Conference \& Exhibition (DATE)}, pages 1303--1306. IEEE, 2020.

\bibitem{chiappetta2016real}
Marco Chiappetta, Erkay Savas, and Cemal Yilmaz.
\newblock Real time detection of cache-based side-channel attacks using hardware performance counters.
\newblock {\em Applied Soft Computing}, 49:1162--1174, 2016.

\bibitem{clavier2007secret}
Christophe Clavier.
\newblock Secret external encodings do not prevent transient fault analysis.
\newblock In {\em Cryptographic Hardware and Embedded Systems-CHES 2007: 9th International Workshop, Vienna, Austria, September 10-13, 2007. Proceedings 9}, pages 181--194. Springer, 2007.

\bibitem{cojocar2020susceptible}
Lucian Cojocar, Jeremie Kim, Minesh Patel, Lillian Tsai, Stefan Saroiu, Alec Wolman, and Onur Mutlu.
\newblock Are we susceptible to rowhammer? an end-to-end methodology for cloud providers.
\newblock In {\em 2020 IEEE Symposium on Security and Privacy (SP)}, pages 712--728. IEEE, 2020.

\bibitem{cojocar2019ecc}
Lucian Cojocar, Kaveh Razavi, Cristiano Giuffrida, and Herbert Bos.
\newblock Exploiting correcting codes: On the effectiveness of {\MakeUppercase{ecc}} memory against rowhammer attacks.
\newblock In {\em 2019 IEEE Symposium on Security and Privacy (SP)}, pages 55--71. IEEE, 2019.

\bibitem{corbet2016kernel}
Jonathan Corbet.
\newblock {\em {\normalfont{Defending against Rowhammer in the kernel}}}, October 2016.
\newblock \url{https://lwn.net/Articles/704920/}.

\bibitem{deridder2021smash}
Finn de~Ridder, Pietro Frigo, Emanuele Vannacci, Herbert Bos, Cristiano Giuffrida, and Kaveh Razavi.
\newblock {SMASH}: Synchronized many-sided rowhammer attacks from {JavaScript}.
\newblock In {\em 30th USENIX Security Symposium (USENIX Security 21)}, pages 1001--1018. USENIX Association, August 2021.

\bibitem{dobraunig2018sifa}
Christoph Dobraunig, Maria Eichlseder, Thomas Korak, Stefan Mangard, Florian Mendel, and Robert Primas.
\newblock Sifa: exploiting ineffective fault inductions on symmetric cryptography.
\newblock {\em IACR Transactions on Cryptographic Hardware and Embedded Systems}, pages 547--572, 2018.

\bibitem{FrodoKemattack}
Michael Fahr, Hunter Kippen, Andrew Kwong, Thinh Dang, Jacob Lichtinger, Dana Dachman-Soled, Daniel Genkin, Alexander Nelson, Ray Perlner, Arkady Yerukhimovich, and Daniel Apon.
\newblock When frodo flips: End-to-end key recovery on frodokem via rowhammer.
\newblock CCS '22, page 979–993, New York, NY, USA, 2022. Association for Computing Machinery.

\bibitem{frigo2020trrespass}
Pietro Frigo, Emanuele Vannacc, Hasan Hassan, Victor Van Der~Veen, Onur Mutlu, Cristiano Giuffrida, Herbert Bos, and Kaveh Razavi.
\newblock {\MakeUppercase{TRR}}espass: Exploiting the many sides of target row refresh.
\newblock In {\em 2020 IEEE Symposium on Security and Privacy (SP)}, pages 747--762. IEEE, 2020.

\bibitem{fuhr2013fault}
Thomas Fuhr, {\'E}liane Jaulmes, Victor Lomn{\'e}, and Adrian Thillard.
\newblock Fault attacks on aes with faulty ciphertexts only.
\newblock In {\em 2013 Workshop on Fault Diagnosis and Tolerance in Cryptography}, pages 108--118. IEEE, 2013.

\bibitem{gierlichs2012infective}
Benedikt Gierlichs, J{\"o}rn-Marc Schmidt, and Michael Tunstall.
\newblock Infective computation and dummy rounds: Fault protection for block ciphers without check-before-output.
\newblock In {\em Progress in Cryptology--LATINCRYPT 2012: 2nd International Conference on Cryptology and Information Security in Latin America, Santiago, Chile, October 7-10, 2012. Proceedings 2}, pages 305--321. Springer, 2012.

\bibitem{gruss2018another}
Daniel Gruss, Moritz Lipp, Michael Schwarz, Daniel Genkin, Jonas Juffinger, Sioli O'Connell, Wolfgang Schoechl, and Yuval Yarom.
\newblock Another flip in the wall of rowhammer defenses.
\newblock In {\em 2018 IEEE Symposium on Security and Privacy (SP)}, pages 245--261. IEEE, 2018.

\bibitem{gruss2016rowhammerjs}
Daniel Gruss, Cl{\'e}mentine Maurice, and Stefan Mangard.
\newblock Rowhammer. js: A remote software-induced fault attack in javascript.
\newblock In {\em International conference on detection of intrusions and malware, and vulnerability assessment}, pages 300--321. Springer, 2016.

\bibitem{gruss2016flush+}
Daniel Gruss, Cl{\'e}mentine Maurice, Klaus Wagner, and Stefan Mangard.
\newblock Flush+ {\MakeUppercase{f}}lush: a fast and stealthy cache attack.
\newblock In {\em International Conference on Detection of Intrusions and Malware, and Vulnerability Assessment}, pages 279--299. Springer, 2016.

\bibitem{heninger2009reconstructing}
Nadia Heninger and Hovav Shacham.
\newblock Reconstructing rsa private keys from random key bits.
\newblock In {\em Annual International Cryptology Conference}, pages 1--17. Springer, 2009.

\bibitem{herath2015these}
Nishad Herath and Anders Fogh.
\newblock These are not your grand {\MakeUppercase{d}}addys cpu performance counters--cpu hardware performance counters for security.
\newblock {\em Black Hat Briefings}, 2015.

\bibitem{irazoqui2016mascat}
Gorka Irazoqui, Thomas Eisenbarth, and Berk Sunar.
\newblock {\MakeUppercase{MASCAT}}: Stopping microarchitectural attacks before execution.
\newblock {\em IACR Cryptol. ePrint Arch.}, 2016:1196, 2016.

\bibitem{islam2019spoiler}
Saad Islam, Ahmad Moghimi, Ida Bruhns, Moritz Krebbel, Berk Gulmezoglu, Thomas Eisenbarth, and Berk Sunar.
\newblock {SPOILER}: Speculative load hazards boost rowhammer and cache attacks.
\newblock In {\em 28th USENIX Security Symposium (USENIX Security 19)}, pages 621--637, Santa Clara, CA, August 2019. USENIX Association.

\bibitem{Islam2022SignatureCA}
Saad Islam, Koksal Mus, Richa Singh, Patrick Schaumont, and Berk Sunar.
\newblock Signature correction attack on dilithium signature scheme.
\newblock {\em 2022 IEEE 7th European Symposium on Security and Privacy (EuroS\&P)}, pages 647--663, 2022.

\bibitem{mus2020quantumhammer}
Saad Islam, Koksal Mus, and Berk Sunar.
\newblock Quantumhammer: A practical hybrid attack on the {\MakeUppercase{luov}} signature scheme.
\newblock In {\em Proceedings of the 2020 ACM SIGSAC Conference on Computer and Communications Security}, pages 1071--1084, 2020.

\bibitem{jattke2022blacksmith}
Patrick Jattke, Victor van~der Veen, Pietro Frigo, Stijn Gunter, and Kaveh Razavi.
\newblock Blacksmith: Scalable rowhammering in the frequency domain.
\newblock In {\em 2022 IEEE Symposium on Security and Privacy (SP)}, volume~1, 2022.

\bibitem{juffinger2023csi}
Jonas Juffinger, Lukas Lamster, Andreas Kogler, Maria Eichlseder, Moritz Lipp, and Daniel Gruss.
\newblock Csi: Rowhammer--cryptographic security and integrity against rowhammer.
\newblock In {\em 2023 IEEE Symposium on Security and Privacy (SP)}, pages 1702--1718. IEEE, 2023.

\bibitem{kangsledgehammer24}
Ingab Kang, Walter Wang, Jason Kim, Stephan van Schaik, Youssef Tobah, Daniel Genkin, Andrew Kwong, and Yuval Yarom.
\newblock Sledgehammer: Amplifying rowhammer via bank-level parallelism.
\newblock In {\em 33rd {USENIX} Security Symposium ({USENIX} Security 24)}, 2024.

\bibitem{kim2014flipping}
Yoongu Kim, Ross Daly, Jeremie Kim, Chris Fallin, Ji~Hye Lee, Donghyuk Lee, Chris Wilkerson, Konrad Lai, and Onur Mutlu.
\newblock Flipping bits in memory without accessing them: An experimental study of dram disturbance errors.
\newblock {\em ACM SIGARCH Computer Architecture News}, 42(3):361--372, 2014.

\bibitem{Kocher2018spectre}
Paul Kocher, Jann Horn, Anders Fogh, , Daniel Genkin, Daniel Gruss, Werner Haas, Mike Hamburg, Moritz Lipp, Stefan Mangard, Thomas Prescher, Michael Schwarz, and Yuval Yarom.
\newblock Spectre attacks: Exploiting speculative execution.
\newblock In {\em 40th IEEE Symposium on Security and Privacy (S\&P'19)}, 2019.

\bibitem{kogler2022halfdouble}
Andreas Kogler, Jonas Juffinger, Salman Qazi, Yoongu Kim, Moritz Lipp, Nicolas Boichat, Eric Shiu, Mattias Nissler, and Daniel Gruss.
\newblock Half-double: Hammering from the next row over.
\newblock In {\em 31st USENIX Security Symposium: USENIX Security'22}, 2022.

\bibitem{DilithimSCA24}
Elisabeth Krahmer, Peter Pessl, Georg Land, and Tim Güneysu.
\newblock Correction fault attacks on randomized crystals-dilithium.
\newblock Cryptology ePrint Archive, Paper 2024/138, 2024.
\newblock \url{https://eprint.iacr.org/2024/138}.

\bibitem{kwong2020rambleed}
Andrew Kwong, Daniel Genkin, Daniel Gruss, and Yuval Yarom.
\newblock {\MakeUppercase{RAMB}}leed: Reading bits in memory without accessing them.
\newblock In {\em 2020 IEEE Symposium on Security and Privacy (SP)}, pages 695--711. IEEE, 2020.

\bibitem{Lipp2018meltdown}
Moritz Lipp, Michael Schwarz, Daniel Gruss, Thomas Prescher, Werner Haas, Anders Fogh, Jann Horn, Stefan Mangard, Paul Kocher, Daniel Genkin, Yuval Yarom, and Mike Hamburg.
\newblock Meltdown: Reading kernel memory from user space.
\newblock In {\em 27th {USENIX} Security Symposium ({USENIX} Security 18)}, 2018.

\bibitem{lipp2020nethammer}
Moritz Lipp, Michael Schwarz, Lukas Raab, Lukas Lamster, Misiker~Tadesse Aga, Cl{\'e}mentine Maurice, and Daniel Gruss.
\newblock Nethammer: Inducing rowhammer faults through network requests.
\newblock In {\em 2020 IEEE European Symposium on Security and Privacy Workshops (EuroS\&PW)}, pages 710--719. IEEE, 2020.

\bibitem{marazzi2023rega}
Michele Marazzi, Flavien Solt, Patrick Jattke, Kubo Takashi, and Kaveh Razavi.
\newblock Rega: Scalable rowhammer mitigation with refresh-generating activations.
\newblock In {\em 44rd IEEE Symposium on Security and Privacy (SP 2023)}. IEEE, 2023.

\bibitem{Molnar2006EnhancingKernel}
Ingo Molnar.
\newblock Enhancing the linux kernel.
\newblock \url{https://lkml.org/lkml/2006/7/25/62}, Jul 2006.
\newblock Accessed: 2024-02-07.

\bibitem{LWERowhammer}
Puja Mondal, Suparna Kundu, Sarani Bhattacharya, Angshuman Karmakar, and Ingrid Verbauwhede.
\newblock A practical key-recovery attack on lwe-based key-encapsulation mechanism schemes using rowhammer.
\newblock {\em 22nd International Conference on Applied Cryptography and Network Security (ACNS)}, 2024.

\bibitem{mus2023jolt}
Koksal Mus, Yarkın Doröz, M.~Caner Tol, Kristi Rahman, and Berk Sunar.
\newblock Jolt: Recovering tls signing keys via rowhammer faults.
\newblock In {\em 2023 IEEE Symposium on Security and Privacy (SP)}, pages 1719--1736, 2023.

\bibitem{quantumhammer}
Koksal Mus, Saad Islam, and Berk Sunar.
\newblock Quantumhammer: A practical hybrid attack on the luov signature scheme.
\newblock In {\em Proceedings of the 2020 ACM SIGSAC Conference on Computer and Communications Security}, CCS '20, page 1071–1084, New York, NY, USA, 2020. Association for Computing Machinery.

\bibitem{mutlu2017rowhammer}
Onur Mutlu.
\newblock The rowhammer problem and other issues we may face as memory becomes denser.
\newblock In {\em Design, Automation \& Test in Europe Conference \& Exhibition (DATE), 2017}, pages 1116--1121. IEEE, 2017.

\bibitem{paterson2012coding}
Kenneth~G Paterson, Antigoni Polychroniadou, and Dale~L Sibborn.
\newblock A coding-theoretic approach to recovering noisy rsa keys.
\newblock In {\em Advances in Cryptology--ASIACRYPT 2012: 18th International Conference on the Theory and Application of Cryptology and Information Security, Beijing, China, December 2-6, 2012. Proceedings 18}, pages 386--403. Springer, 2012.

\bibitem{payer2016hexpads}
Mathias Payer.
\newblock Hex{\MakeUppercase{pads}}: a platform to detect “stealth” attacks.
\newblock In {\em International Symposium on Engineering Secure Software and Systems}, pages 138--154. Springer, 2016.

\bibitem{pessl2016drama}
Peter Pessl, Daniel Gruss, Cl{\'e}mentine Maurice, Michael Schwarz, and Stefan Mangard.
\newblock {DRAMA}: Exploiting {DRAM} addressing for {Cross-CPU} attacks.
\newblock In {\em 25th USENIX Security Symposium (USENIX Security 16)}, pages 565--581, Austin, TX, August 2016. USENIX Association.

\bibitem{islam2022dilithium}
Islam Saad, Koksal Mus, Richa Singh, Patrick Schaumont, and Berk Sunar.
\newblock A signature correction attack on the post-quantum scheme dilithium.
\newblock In {\em Proceedings of the IEEE European Workshop on Security and \& Privacy}, 2022.

\bibitem{seaborn2015exploiting}
Mark Seaborn and Thomas Dullien.
\newblock Exploiting the dram rowhammer bug to gain kernel privileges.
\newblock {\em Black Hat}, 15:71, 2015.

\bibitem{tatar2018hammertime}
Andrei Tatar, Cristiano Giuffrida, Herbert Bos, and Kaveh Razavi.
\newblock Defeating software mitigations against rowhammer: A surgical precision hammer.
\newblock In Michael Bailey, Thorsten Holz, Manolis Stamatogiannakis, and Sotiris Ioannidis, editors, {\em Research in Attacks, Intrusions, and Defenses}, pages 47--66, Cham, 2018. Springer International Publishing.

\bibitem{tatar2018throwhammer}
Andrei Tatar, Radhesh~Krishnan Konoth, Elias Athanasopoulos, Cristiano Giuffrida, Herbert Bos, and Kaveh Razavi.
\newblock Throwhammer: Rowhammer attacks over the network and defenses.
\newblock In {\em 2018 USENIX Annual Technical Conference (USENIX ATC 18)}, pages 213--226, Boston, MA, July 2018. USENIX Association.

\bibitem{tobah2022spechammer}
Youssef Tobah, Andrew Kwong, Ingab Kang, Daniel Genkin, and Kang~G Shin.
\newblock Spechammer: Combining spectre and rowhammer for new speculative attacks.
\newblock In {\em 2022 IEEE Symposium on Security and Privacy (SP)}, pages 681--698. IEEE, 2022.

\bibitem{tupsamudre2014destroying}
Harshal Tupsamudre, Shikha Bisht, and Debdeep Mukhopadhyay.
\newblock Destroying fault invariant with randomization: A countermeasure for aes against differential fault attacks.
\newblock In {\em Cryptographic Hardware and Embedded Systems--CHES 2014: 16th International Workshop, Busan, South Korea, September 23-26, 2014. Proceedings 16}, pages 93--111. Springer, 2014.

\bibitem{vanbulck2020lvi}
Jo~Van~Bulck, Daniel Moghimi, Michael Schwarz, Moritz Lipp, Marina Minkin, Daniel Genkin, Yarom Yuval, Berk Sunar, Daniel Gruss, and Frank Piessens.
\newblock {LVI: Hijacking Transient Execution through Microarchitectural Load Value Injection}.
\newblock In {\em 41th IEEE Symposium on Security and Privacy (S\&P'20)}, 2020.

\bibitem{vandeVen2005RandomizeStackPointer}
Arjan van~de Ven.
\newblock Patch 4/6 randomize the stack pointer.
\newblock {\em LWN.net}, Jan 2005.
\newblock Accessed: 2024-02-07.

\bibitem{van2016drammer}
Victor Van Der~Veen, Yanick Fratantonio, Martina Lindorfer, Daniel Gruss, Cl{\'e}mentine Maurice, Giovanni Vigna, Herbert Bos, Kaveh Razavi, and Cristiano Giuffrida.
\newblock Drammer: Deterministic rowhammer attacks on mobile platforms.
\newblock In {\em Proceedings of the 2016 ACM SIGSAC conference on computer and communications security}, pages 1675--1689, 2016.

\bibitem{vanschaik2019ridl}
Stephan van Schaik, Alyssa Milburn, Sebastian Österlund, Pietro Frigo, Giorgi Maisuradze, Kaveh Razavi, Herbert Bos, and Cristiano Giuffrida.
\newblock {RIDL}: Rogue in-flight data load.
\newblock In {\em S\&{P}}, May 2019.

\bibitem{wang2021discreetpara}
Z.~Wang, W.~Liu, and Y.~Wang.
\newblock Discreet-para: Rowhammer defense with low cost and high efficiency.
\newblock In {\em 2021 IEEE 39th International Conference on Computer Design (ICCD)}, pages 1--8. IEEE, 2021.

\bibitem{weissman2019jackhammer}
Zane Weissman, Thore Tiemann, Daniel Moghimi, Evan Custodio, Thomas Eisenbarth, and Berk Sunar.
\newblock Jackhammer: Efficient rowhammer on heterogeneous fpga-cpu platforms.
\newblock {\em IACR Transactions on Cryptographic Hardware and Embedded Systems}, 2020(3):169–195, Jun. 2020.

\bibitem{wolfssl}
wolfSSL Inc.
\newblock wolfssl (5.5.0), Aug 2022.

\bibitem{xiao2016one}
Yuan Xiao, Xiaokuan Zhang, Yinqian Zhang, and Radu Teodorescu.
\newblock One bit flips, one cloud flops: {Cross-VM} row hammer attacks and privilege escalation.
\newblock In {\em 25th USENIX Security Symposium (USENIX Security 16)}, pages 19--35, Austin, TX, August 2016. USENIX Association.

\bibitem{HiRA2022}
A.~Giray Yağlikçi, Ataberk Olgun, Minesh Patel, Haocong Luo, Hasan Hassan, Lois Orosa, Oğuz Ergin, and Onur Mutlu.
\newblock Hira: Hidden row activation for reducing refresh latency of off-the-shelf dram chips.
\newblock In {\em 2022 55th IEEE/ACM International Symposium on Microarchitecture (MICRO)}, pages 815--834, 2022.

\bibitem{yen2000checking}
Sung-Ming Yen and Marc Joye.
\newblock Checking before output may not be enough against fault-based cryptanalysis.
\newblock {\em IEEE Transactions on computers}, 49(9):967--970, 2000.

\bibitem{zhang2016cloudradar}
Tianwei Zhang, Yinqian Zhang, and Ruby~B Lee.
\newblock Cloudradar: A real-time side-channel attack detection system in clouds.
\newblock In {\em International Symposium on Research in Attacks, Intrusions, and Defenses}, pages 118--140. Springer, 2016.

\end{thebibliography}


\vspace{12pt}

\end{document}